%% file: AnnRev-LFU_arXiv_v1.tex
\documentclass{ar-1col-S2O}

\usepackage{bm,amssymb,slashed,graphicx,multirow,soul,mathtools,xspace,array}
\usepackage[inline]{enumitem}
\usepackage{makecell}
\usepackage{placeins}
\usepackage{xcolor}
\usepackage{needspace}
\usepackage{threeparttable} 
\usepackage{mciteplus}
 \usepackage[normalem]{ulem}
 
\definecolor{nicered}{rgb}{0.7,0.1,0.1}
\definecolor{nicegreen}{rgb}{0.1,0.5,0.1}
\definecolor{niceblue}{rgb}{0.1,0.1,0.5}

\usepackage[bookmarks=false]{hyperref}
\hypersetup{colorlinks,citecolor= nicegreen,linkcolor= niceblue,urlcolor= niceblue}

\newcommand{\nn}{\nonumber}

\newcommand{\TeV}{\text{TeV}}
\newcommand{\Leff}{\Lambda_{\text{eff}}}
\newcommand{\g}{\gamma}
\newcommand{\cbar}{\bar{c}}

\newcommand{\Qbar}{\bar{Q}}
\newcommand{\Lbar}{\bar{L}}
\newcommand{\Ubar}{\bar{u}}

\newcommand{\ebar}{\bar{e}}

\newcommand{\NR}{N_{\!R}}
\newcommand{\op}[1]{\mathcal{O}_{\text{#1}}}
\newcommand{\Op}[1]{O_{\text{#1}}}
\newcommand{\wc}[1]{c_{\text{#1}}}
\newcommand{\wcb}[1]{\bar{c}_{\text{#1}}}
\newcommand{\Wc}[1]{C_{\text{#1}}}
\newcommand{\opn}[1]{\mathcal{O}^{\prime}_{\text{#1}}}
\newcommand{\Opn}[1]{O^{\prime}_{\text{#1}}}
\newcommand{\msqmiss}{m^2_{\text{miss}}}
\newcommand{\pstarell}{|\vec p^{\,*}_\ell|}

\newcommand{\taubar}{\bar{\tau}}
\newcommand{\Bbar}{\bar{B}}
\newcommand{\Dbar}{\bar{D}}
\newcommand{\Dstar}{D^*}
\newcommand{\Dx}{D^{(*)}}

\newcommand{\Hb}{H_b}
\newcommand{\Hc}{H_c}
\newcommand{\Hu}{H_u}
\newcommand{\Hbbar}{\bar{H}_b}
\newcommand{\Hcbar}{\bar{H}_c}

\newcommand{\Rx}[1]{R(#1)}
\newcommand{\RD}{\Rx{D}}
\newcommand{\RDs}{\Rx{\Dstar}}
\newcommand{\RDx}{\Rx{\Dx}}
\newcommand{\RJ}{\Rx{J/\psi}}
\newcommand{\RLc}{\Rx{\Lambda_c}}

\newcommand{\bctn}{b \to c \tau \bar{\nu}_\tau}
\newcommand{\bcelln}{b \to c \ell \bar{\nu}_\ell}
\newcommand{\bcln}{b \to c l \bar{\nu}_l}
\newcommand{\butn}{b \to u \tau \bar{\nu}_\tau}
\newcommand{\bsellell}{b \to s \ell \bar{\ell}}
\newcommand{\bsmumu}{b \to s \mu \bar{\mu}}
\newcommand{\bsnunu}{b \to s \nu \bar{\nu}}
\newcommand{\Lqcd}{\Lambda_{\text{QCD}}}

\newcommand{\matrixelem}[3]{\langle #1 | #2 | #3 \rangle}
\newcommand{\MatrixElem}[3]{\big\langle #1 \big| #2 \big| #3 \big\rangle}
\newcommand{\BR}[1]{\text{Br}[#1]}

\makeatletter
\g@addto@macro\bfseries{\boldmath}
\makeatother

\allowdisplaybreaks
\begin{document}
%Fix citation format: square brackets
%Fix equation format: (), no period
%Fix link colors
\makeatletter
	\def\@cite#1#2{[{#1\if@tempswa , #2\fi}]}%
	\def\@eqnnum{{\normalfont \normalcolor (\theequation)}}
	\def\tagform@#1{\maketag@@@{\ignorespaces(#1)\unskip}}
	\def\@linkcolor{niceblue}
\makeatother

%\title{Status and strategies for the observation of new physics in semitauonic $b$-hadron decays}
%\title{Discovering new physics with semitauonic $b$-hadron decays: status and strategies}
\title{Discovering new physics with semitauonic $b$-hadron decays}

%%Authors, affiliations address.
\author{Markus Prim$^1$ and Dean J. Robinson$^{2,3}$
\affil{$^1$Physikalisches Institut der Rheinischen Friedrich-Wilhelms-Universit\"at Bonn, 53115 Bonn, Germany}
\affil{$^2$Ernest Orlando Lawrence Berkeley National Laboratory, University of California, Berkeley, CA 94720, USA}
\affil{$^3$Leinweber Institute for Theoretical Physics at UC Berkeley, University of California, Berkeley, CA 94720, USA}}
%
%%Abstract
\begin{abstract}
Anomalous lepton flavor universality violation (LFUV) observed in semileptonic decays of beauty hadrons may be the first signal of new physics (NP) beyond the Standard Model.
We review the essential theoretical concepts and experimental considerations 
required for self-consistent and unbiased NP interpretations of these LFUV anomalies.
We examine the status of present-day analyses, 
and assess the core challenges to the future precision studies that will be required to conclusively establish or reject an NP signal, 
as well as promising strategies to resolve them.
\end{abstract}

%%Keywords, etc.
%\begin{keywords}
%lepton flavor universality violation. semileptonic decays, new physics
%keywords, separated by comma, no full stop, lowercase
%\end{keywords}
\maketitle

\tableofcontents

\section{Introduction}
Mid-decade in the 2020s, particle physicists remain intrigued
by a persistent puzzle in lepton flavor universality violation (LFUV) involving semileptonic decays of beauty hadrons to charm and $\tau$ leptons.
In particular, $\bctn$ decay rates measured in various processes and experiments over the last fifteen years
appear to be consistently larger than predicted by the Standard Model (SM), at more than $3\sigma$ significance.

In the SM, amplitudes to electron, muon, or $\tau$ final states should be universal up to universality-violating effects proportional to the lepton mass.
The large $\tau$ mass means such effects can be sizable. 
Semileptonic decays provide a precision laboratory to test these effects not only because of their abundant statistics,
but also because of well-developed theoretical frameworks that permit prediction of LFUV effects to at least several percent or better,
whether for the SM or also including New Physics (NP).
Comparing $\bcelln$ ($\ell = e$ or $\mu$) and $\bctn$ decays allows one, in principle, to perform two types of inference: 
first, construct hypothesis tests of the SM itself, and second, investigate compatibility of any anomalous results with alternate NP explanations.

Since the large excess first reported by BaBar in $\Bbar \to \Dx \tau \nu$ decays,
subsequent higher precision measurements from Belle, Belle~II, and LHCb 
over a range of observables and decay modes have reduced the size of the putative LFUV anomalies,
while their combined statistical significance versus the SM has nonetheless remained stable above $3\sigma$.
The state of play is a race between a shrinking NP signal and its increasing precision. 
In one outcome the signal may converge back to SM predictions;
in another it may established as a highly significant but small deviation from the SM.
Distinguishing these scenarios requires precision control of the measurements and predictions.
In practice, however, various systematic limitations and biases of both theoretical and experimental origin will pose deep challenges 
to realizing the full statistical power of future datasets, 
that may be required to establish or reject LFUV anomalies as NP signals.
This review is devoted to providing a comprehensive survey of the theoretical concepts and experimental considerations 
required for self-consistent and unbiased NP interpretations of the LFUV anomalies;
the challenges to present-day and future precision analyses, as well as promising strategies to resolve them.

On the theoretical side, ultraviolet (UV) motivations may be critical to resolving degeneracies in model-independent searches over high-dimensional spaces of NP couplings.
We outlay the main elements of matching of low-scale weak effective theory operators onto electroweak-consistent effective field theories, 
and categorization of simplified models that may generate the required charged currents. 
We survey plausible UV complete flavor model frameworks capable of generating these anomalies,
and the space of applicable precision flavor constraints,
including an assessment of mechanisms and frameworks that should be revisited in light 
of the now-defunct $\bsellell$ LFUV anomalies 
(up to less theoretically-clean putative tensions with $\bsmumu$ branching fractions) 
and the now-disfavored tension in the muon anomalous magnetic moment, as well as new possible tensions in $\bsnunu$. 

In parallel, precision NP interpretation of semileptonic data relies on exquisite theoretical control of hadronic matrix elements. 
We highlight the essential elements required to understand state-of-the-art theoretical frameworks for both exclusive and inclusive processes, 
including both continuum and lattice QCD (LQCD)-based formalisms,
and resolve important subtleties and confusions in form factor descriptions. 
For exclusive processes, we summarize and compare the various advantages, limitations, and caveats on each theoretical approach,
how they are combined into present day precision descriptions, 
and what challenges must be resolved to enable future precision analysis.
We further discuss the power of fully-differential probes, and the construction of optimized observables for NP discovery.

On the experimental side, 
we review the latest LFUV measurements in the last five years, 
as well as the progress in control of the dominant backgrounds that limit them,
especially descriptions of the orbitally excited $D^{**}$ modes.
We then examine caveats in the interpretations of global averages, 
not only in the averaging procedures themselves, 
but also the impact of biases from Monte Carlo template mismatches in fits to data,
and the effects of shared hadronic systematic uncertainties.
We illustrate a new class of future sizable systematic uncertainties, 
that arise from degeneracies between hadronic form factor parametrizations and NP couplings.
Finally we comment on the outlook for a rich experimental program, including $B_s$--$\bar{B}_s$ mixing, 
$\bsnunu$ and $\Upsilon$ decays and high-$p_T$ measurements, 
that will be required to conclusively interpret NP phenomenology.

\section{New Physics in semileptonic decays}
\label{sec:npsl}
\subsection{Low-energy NP operators}
Below the electroweak (EW) scale, 
such that the top quark, the Higgs, and the $W$ and $Z$ bosons are integrated out,
the weak EFT~(WET) contains the interaction terms $(1/\Lambda_{\text{eff}}^2) \sum_X \wc{X} \op{X}$ 
in which the $\op{X}$ are four-fermion operators that mediate $\bctn$ decays.
The $\wc{X}$ are Wilson coefficients (WCs), and in general may be complex. 
They are normalized versus the SM by setting the effective scale
\begin{equation}
\label{eqn:Leff}
	\Leff =\big(2\sqrt 2 G_F |V_{cb}|\big)^{-1/2}\!\simeq 0.87 \Big[ 40 \!\times\! 10^{-3} / |V_{cb}|\Big]^{1/2}\,\TeV\,.
\end{equation}
The purported $\mathcal{O}(20\%)$ upwards deviation in a $\bctn$ rate implies, assuming interference with the SM,
a mediator mass scale $\sim \Leff/\sqrt{0.1} \sim \text{few}\,\TeV$.
The $\op{X}$ have the basis
\begin{align}
	\op{SR}	& = \big( \cbar_L b_R \big) \big( \taubar_R \nu_L \big)\,, 			&\op{SL}  & = \big( \cbar_R b_L \big) \big( \taubar_R \nu_L \big)\,,\label{eqn:opsL} \\
	\op{VR}	& = \big(\cbar_R \g^\mu  b_R\big)\big(\taubar_L \g_\mu \nu_L \big)\,, {\kern-2ex}&\op{VL}  & = \big(\cbar_L \g^\mu  b_L\big)\big(\taubar_L \g_\mu \nu_L \big)\,, &
	\op{T}  & = \big(\cbar_R \sigma^{\mu\nu} b_L \big)\big(\taubar_R \sigma_{\mu\nu} \nu_L\big)\,,  \nn
\end{align}
with $\sigma^{\mu\nu} = (i/2)[\g^\mu, \g^\nu]$. 
The axial-tensor operator $\big(\cbar_R \sigma^{\mu\nu} \g^5 b_L \big)\big(\taubar_R \sigma_{\mu\nu} \g^5 \nu_L\big)$ is related to the tensor operator by a duality identity
(see Eq.~\eqref{eqn:sigdual} below) which also ensures $\big(\cbar_L \sigma^{\mu\nu} b_R \big)\big(\taubar_R \sigma_{\mu\nu} \nu_L\big)$ vanishes.
Operators with three or more Dirac matrices are reduced to this basis by the Chisholm identity.
In some contexts one instead uses scalar and pseudoscalar operators $\op{S,P} \equiv \op{SR} \pm \op{SL}$ and vector and axial-vector operators $\op{V,A} \equiv \op{VR} \pm \op{VL}$.
With a right-handed sterile neutrino, $\NR$, in addition one may have
\begin{align}
	\opn{SR}	& = \big( \cbar_L b_R \big) \big( \taubar_L \NR \big)\,, 			&\opn{SL}  & = \big( \cbar_R b_L \big) \big( \taubar_L \NR \big)\,,  \label{eqn:opsR} \\
	\opn{VR}	& = \big(\cbar_R \g^\mu  b_R\big)\big(\taubar_R \g_\mu \NR \big)\,, {\kern-2ex}&\opn{VL}  & = \big(\cbar_L \g^\mu  b_L\big)\big(\taubar_R \g_\mu \NR \big)\,, &
	\opn{T}  & = \big(\cbar_L \sigma^{\mu\nu} b_R \big)\big(\taubar_L \sigma_{\mu\nu} \NR \big)\,. \nn
\end{align}
One may consider $\butn$, replacing the charm with the up quark, though we do not focus on these in this review.
In light semileptonic decays, $\ell = \mu$ or $e$ replaces the $\tau$.

The SM (at leading order in $G_F$) corresponds to $\wc{VL} =1$ with all others zero.
These ten operators fully describe the space of NP interactions in the low energy WET.~\footnote{%
A left-handed sterile neutrino, $N_L$, would imply a copy of the $\op{X}$ basis with $\nu_L \to N_L$.
Matched onto SMEFT, such operators enter at dimension seven or higher, and thus are power suppressed.}~\footnote{%
The WCs $\wcb{X}$ for the conjugate operators of form $\bar{O}_{\text{X}} = (\bar{b} \Gamma c)(\bar{\nu}_L \Gamma^\prime \tau)$
are related to the $\wc{X}$ via $\wcb{VL,VR} =\wc{VL,VR}^*$ and $\wcb{SL,SR,T} = -\wc{SR,SL,T}^*$.}
In contrast to WCs of a UV theory (see below), the WET WCs are those that are directly constrainable from $b$-hadron decay measurements.
Identifying an NP WET WC different from zero at high significance would establish detection of new physics.

\subsection{SMEFT matching and simplified models}
\label{sec:leptoquarks}
The interpretation of measurements or limits on WET WCs in terms of EW-consistent effective field theories (EFTs) and related simplified models
is crucial not only because it reveals complementary constraints from other collider probes and flavor data,
but also because UV completions must themselves be EW consistent and map onto these models.
Thus these interpretations illustrate which combinations of WET WCs might be well motivated from a UV perspective.
This provides targeted simplified model hypotheses for which it may be more feasible to recover convergent CLs versus fits over the full WC space.

The renormalization group (RG) evolution between scale $\mu < m_b$ and the EW scale is dominated by QCD effects,
while higher-order EW RG effects and associated mixing are negligible; see e.g. Ref.~\cite{Gonzalez-Alonso:2017iyc}.
Under QCD RG evolution the vector currents do not renormalize, while the scalar and tensor currents do.
In the leading-log approximation the running for the WCs from $m_t$ down to $\mu$ 
for $X = \text{SR},\text{SL},\text{T}$ at one-loop order~\cite{Dorsner:2013tla,Freytsis:2015qca}
\def\gX{ \g_{\text{X}_{\vphantom{0}}} }
\begin{align}
\wc{X}(\mu)   
	& = \Big[ \alpha_s(m_b)/ \alpha_s(\mu)\Big]^{\frac{\gX}{2\beta_0^{(4)}}} \Big[\alpha_s(m_t)/\alpha_s(m_b)\Big]^{\frac{\gX}{2\beta_0^{(5)}}} \wc{X}(m_t)
	 \equiv \rho_{\text{X}}(\mu; m_t) \wc{X}(m_t)\,, \label{eqn:qcdrge}
\end{align}
with anomalous dimensions $\g_{\text{SR},\text{SL}} = -8$, $\g_{\text{T}} = 8/3$ 
and the one-loop $\beta$-function coefficient $\beta^{(n)}_0 = 11 - 2 n /3$ for $n$ flavors.
Choosing $\mu = \sqrt{m_c m_b}$---a convenient matching scale for QCD onto heavy quark effective theory (HQET)---one
finds $\rho_{\text{SR},\text{SL}} \simeq 1.6$ and $\rho_{\text{T}} \simeq 0.85$,
using the four-loop running for $\alpha_s$ itself~\cite{Czakon:2004bu,VANRITBERGEN1997379}.
A consistent treatment to three loops~\cite{Gonzalez-Alonso:2017iyc} (for real WCs) results in further few percent corrections.
(The matching scale for $\butn$ decays might be chosen differently, and may include additional thresholds.)

Assuming linearly-realized EW symmetry breaking (cf. nonlinear realizations~\cite{Buchalla:2012qq,Cata:2015lta} and references therein)
at the EW scale the renormalized WET operators match onto Standard Model EFT (SMEFT) operators in $\mathcal{L}_{\text{SMEFT}} \supset 1/\Leff^2 \,\sum_C \Wc{X} \Op{X} + \ldots$.
At dimension six, omitting all flavor indices, a common basis choice is
\begin{align}
	\Op{SR} & = \big(\Qbar_L^a d_R\big) \big(\ebar_R L_L^a\big)\,, & 
		\Op{SL} & = \epsilon_{ab} \big(\Ubar_R  Q_L^a\big) \big(\ebar_R L_L^b\big)\,, \nn\\
	\Op{VL} & =  \big(\Qbar_L^a \g^\mu T^A_{ab} Q_L^b \big) \big(\Lbar_L^c \g_\mu T^A_{cd} L_L^d \big)\,, {\kern-10ex} &
		\Op{T} & = \epsilon_{ab} \big(\Ubar_R \sigma^{\mu\nu} Q_L^a\big) \big(\ebar_R \sigma_{\mu\nu}  L_L^b\big)\,, {\kern-10ex} \label{eqn:OpsL}
\end{align}
in which $a,b,\ldots$ are $SU(2)_L$ fundamental indices and $T^A$ are the $SU(2)_L$ generators.
Note at tree level $\op{VR}$ matches onto the dimension-8 operator $\Op{VR} = (\Ubar_R \g^\mu d_R)\epsilon_{bc}(\Lbar_L^a H^a \g^\mu H^b L_L^c)$,
that is power suppressed by $v^2/\Leff^2$.
Contributions to $\Wc{VR(L)}$ may also arise from the dimension-6 operator 
$\epsilon_{ab}(H^a D_\mu H^b) \Ubar_{R(L)} \g^\mu d_{R(L)}$ composed with the SM charged current coupling to leptons~\cite{Cirigliano:2009wk,Cata:2015lta}.
These are still power suppressed, but they are also necessarily lepton-flavor universal and thus tightly constrained by measurements of $\bcln$.

The choice of basis in Eq.~\eqref{eqn:OpsL} coincides with the Warsaw basis~\cite{Grzadkowski:2010es} 
in which these operators would be respectively written as $O_{ledq}$, $O_{lequ}$, $O^{(3)}_{lq}$, $O^{(3)}_{lequ}$.
We use notation that renders the matching more transparent; 
care should also be taken with regard to choice of effective scale in comparing results throughout the literature.
In addition to the operators in Eq.~\eqref{eqn:opsL}, the basis~\eqref{eqn:OpsL} also produces other four-fermion operators, 
such as those mediating $\bsnunu$,
generating related collider and flavor constraints.
If one includes a right-handed neutrino in the SMEFT, then the operators in Eq.~\eqref{eqn:opsR} match onto
\begin{align}
	\Opn{SR} & = \epsilon_{ab}\big(\Qbar_L^a  d_R\big)\big(\Lbar_L^b \NR \big)\,, &  
		\Opn{SL}&=\big(\Ubar_R  Q_L^a \big) \big(\Lbar_L^a  \NR \big)\,,\nn \\
	\Opn{VR} & =\big(\Ubar_R \g^\mu  d_R\big)\big(\ebar_R \g_\mu \NR \big)\,, {\kern-15pt} &
		\Opn{T} & = \epsilon_{ab} \big(\Qbar_L^a \sigma^{\mu\nu} d_R\big)\big(\Lbar_L^b\sigma_{\mu\nu} \NR \big)\,,{\kern-22pt} \label{eqn:OpsR}
\end{align}
while $\opn{VL}$ matches onto the dimension-8 operator $\Opn{VL}=\epsilon_{ab}(\Qbar^a_L \tilde H^{\dagger b} \g^\mu H^{\dagger c} Q^c_L \big)\big(\ebar_R \g_\mu \NR \big)$.

Above the EW scale, simplified models with mediator mass scale $M \gtrsim \Leff$ may generate the SMEFT operators $\Op{X}$.
EW RG effects in the running from $m_t$ up to $M$ may, however, induce non-negligible running 
and mixing on top of the QCD RG evolution~\cite{Gonzalez-Alonso:2017iyc}. 
In particular, $\mathcal{O}(20\%)$ mixing effects may be induced between $\Wc{SL}$ and $\Wc{T}$ in running up to the $\TeV$ scale. 
(That sizable EW-induced mixing may occur between these two operators is made clearer in the Warsaw notation:  $O_{lequ}$ and  $O^{(3)}_{lequ}$, respectively.)
While a full treatment of such effects is beyond the scope of this review, they should be kept in mind 
when considering which subsets of WET WCs are motivated by any particular simplified model or UV completion.

\begin{figure}[t]
	\centering
	\begin{minipage}{0.49\textwidth}
        \centering
	\includegraphics[width = \linewidth]{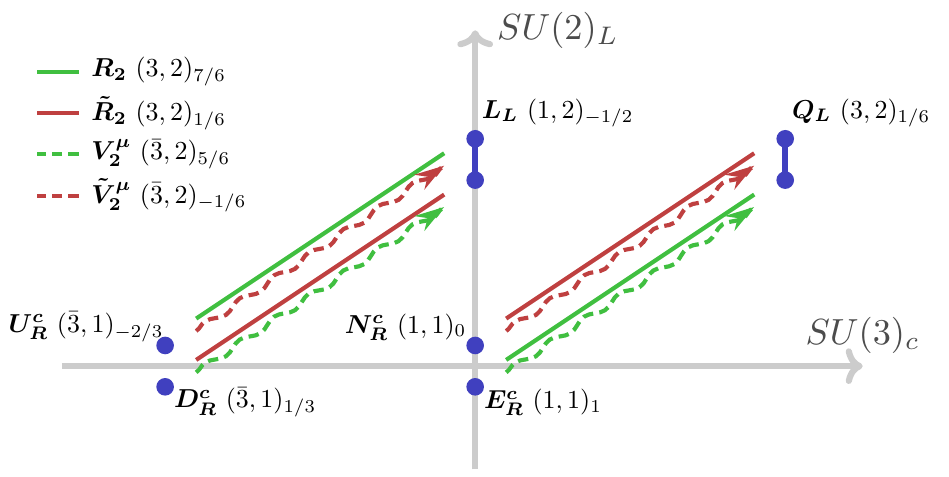} 
	\end{minipage}\hfill%
	\begin{minipage}{0.49\textwidth}
        \centering
	\includegraphics[width = \linewidth]{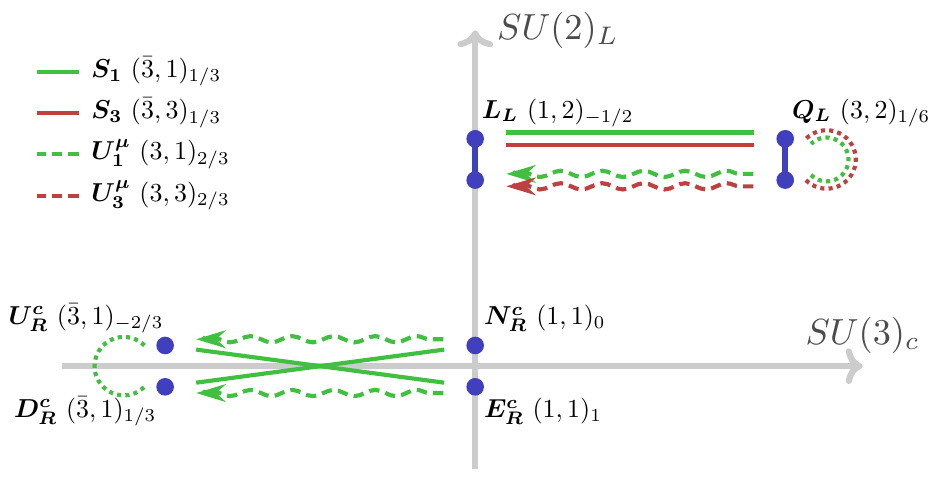} 
	\end{minipage}\hfill%
	\begin{minipage}{0.49\textwidth}
	\centering
	\includegraphics[width =\linewidth]{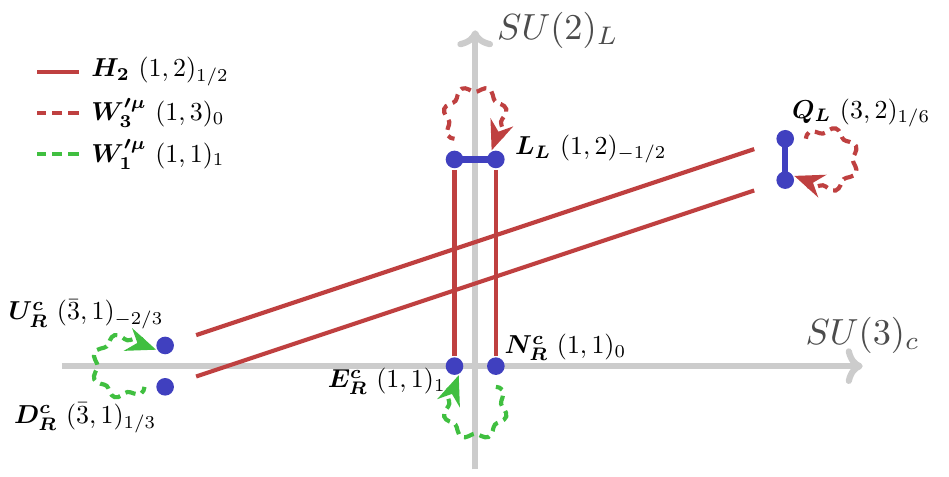}%
	\vspace{2pt}
	\end{minipage}%
	\begin{minipage}{0.55\textwidth}
	\newcolumntype{C}{ >{\raggedright\arraybackslash $} m{2.25cm} <{$}}
	\newcolumntype{D}{ >{\raggedright\arraybackslash $} m{5.25cm} <{$}}
	\centering
	\scalebox{0.7}{
	\begin{tabular}{CD}
	\hline
	\text{Mediator} & \text{WET operators, $r\simeq 1.9$} \\
	\hline\hline
	\bm{R_2}(3,2)_{7/6} & 4r \op{SL} + \op{T} \\
	\bm{\tilde{R}_2}(3,2)_{1/6} & 4r \opn{SR} + \opn{T} \\
	\bm{V_2}^\mu(\bar{3},2)_{5/6} & \op{SR}\\
	\bm{\tilde{V}_2}^\mu(\bar{3},2)_{-1/6} & \opn{SL}\\
	\hline
	\bm{S}_1(\bar{3},1)_{1/3} & \op{VL}, 4r \op{SL} - \op{T}, \opn{VR}, 4r \opn{SR} - \opn{T}\\
	\bm{S}_3(\bar{3},3)_{1/3} & \op{VL}\\
	\bm{U_1}^\mu(3,1)_{2/3} & \op{SR}, \op{VL},  \opn{SL}, \opn{VR}\\
	\bm{U_3}^\mu(3,3)_{2/3} & \op{VL}\\
	\hline
	\bm{H_2}(1,2)_{1/2} & \op{SR}, \op{SL}, \opn{SR}, \opn{SL}\\
	\bm{W_3}^{\prime\mu}(1,3)_0 & \op{VL}\\
	\bm{W_1}^{\prime\mu}(1,1)_1 & \opn{VR}\\
	\hline
	\end{tabular}
	}\vspace{2pt}
	\end{minipage}
	\caption{
	Representation of couplings of EW doublet leptoquarks (top left), EW singlet plus triplet leptoquarks (top right) and scalar and vector bosons (bottom)
	to SM fermions plus $\NR$. 
	The operators (up to conjugation) can be read off the figure: solid lines (dashed wavy arrows) indicate scalar (vector) mediators,
	with the vector arrow from field $X$ to $Y$ indicating a current $\bar{X} \g^\mu Y$.
	Only simplified model mediators capable of generating semileptonic decays are shown,
	and flavor indices are omitted. Dotted lines indicated scalar LQ operators that do not generate semileptonic decays.
	The leptoquark nomenclature follows the notation of Ref.~\cite{Dorsner:2016wpm}.}
	\label{fig:smpmodels}
\end{figure}

Discussions of possible simplified models 
generating LFUV anomalies through the SMEFT operators~\eqref{eqn:OpsL} or~\eqref{eqn:OpsR}
are found extensively throughout the literature:
the wide range of implicated collider and flavor bounds means that most publications focus on one particular simplified setup (or UV completion).
For this review, 
we categorize the simplified model mediators capable of generating semileptonic decays, 
and their couplings, through a novel visual representation shown in Fig.~\ref{fig:smpmodels}.
This categorization maps the SMEFT fermion fields onto weights in the $SU(3)_c \times SU(2)_L$ plane, 
with an additional hypercharge scalar weight. 
We also include mediators coupling to $\NR$.
We use the notation of Ref.~\cite{Dorsner:2016wpm} for the leptoquarks;
the notation for the heavy Higgs-like doublet, $H_2$ and $W^{\prime}$ is chosen to match.
The effective couplings to the WET operators of Eqs~\eqref{eqn:opsL} and~\eqref{eqn:opsR} are also provided in Fig.~\ref{fig:smpmodels}, 
with $r \equiv \rho_{\text{SR}, \text{SL}}/\rho_{\text{T}} \simeq 1.9$ per Eq.~\eqref{eqn:qcdrge}.
In the case of $S_1$, the four WET operators are generated by three UV operators, and therefore their couplings are linearly dependent.

\subsection{Flavor and the ultraviolet}
The simplified models above are not UV completions.
Additional dynamics at or above the mediator scale $M$ is required to (naturally) generate the simplified theory.
Tight bounds from precision measurements of $\bcelln$ decays ($\ell = e$, $\mu$), 
that are closely compatible with SM predictions,
require that the simplified model mediators couple predominantly to the second and third SM generations.
This nontrivial flavor structure---flavor 
anarchy is strongly excluded~\cite{Carpentier:2010ue}---must also be inherited by any UV completion.

Because the NP effective scale $\Leff \sim \text{TeV}$, in a flavor-compatible UV model the couplings of NP to the first generation must be suppressed
to avoid tight bounds from: \begin{enumerate*}[label = \bf(\roman*)] 
	\item kaon mixing;
	\item charged lepton flavor violation;
	\item pure leptonic decays, and rare meson decays;
	\item collider bounds on leptoquark decays to quark-lepton final states, and distortions of high $p_T$ tails~\cite{Greljo:2018tzh};
	\item higher-order corrections to EW precision observables, $Z$ and $W$ decays, and Higgs processes;
	\item atomic parity violation~\cite{Suzuki:1997ma} and electric dipole moments;
	\item neutrino masses and mixings;
	\item the $B_c$ lifetime~\cite{Li:2016vvp,Alonso:2016oyd}.
\end{enumerate*}
In the context of leptoquarks, a review of many of these constraints can be found in Refs~\cite{Dorsner:2016wpm, Schmaltz:2018nls}.
The high $p_T$ tails from Run 2 LHC data already constrain the scalar, vector, and tensor WCs at the $0.3$, $0.6$ and $0.15$ level respectively~\cite{Greljo:2018tzh}, 
probing each down by a further factor of 3 at the end of the HL-LHC.
Constraints on the $B_c$ lifetime have particular impact on $\wc{P}$, because its current induces a chiral flip resulting in a relative $m_b/m_\tau$ enhancement versus the SM $B_c \to \tau \nu$ amplitude.
The current $\mathcal{O}(20\%)$ size of the LFUV anomaly naively disfavors the $H_2$ simplified model, as well as other pure scalar-coupled leptoquarks.
Including effects of template mismatch biases could alter this conclusion, however (see Sec.~\ref{sec:worldav}).

While the literature studying flavor-compatible UV models that explain the LFUV anomalies is extensive,
most literature through 2022 sought to explain also the $R_K$ and $R_{K^*}$ LFUV anomalies, 
now obsolete~\cite{LHCb:2022qnv,LHCb:2022vje},
and/or the now-disfavored muon $g-2$ anomaly. 
See e.g. Ref.~\cite{Buttazzo:2017ixm} for a summary of combined LFUV explanations.
This leads to a present-day dichotomy: 
approaches whose viability is improved by removing the requirement to also explain $R_K$ and $R_{K^*}$ or muon $g-2$;
or approaches which necessarily predict deviations in either of the latter, and are thus now disfavored.
Moreover, new possible anomalies in $\bsnunu$ should be considered~\cite{Belle-II:2023esi},
and putative tensions~\cite{DeFazio:2026gju} in theoretical predictions for $\bsmumu$ branching fractions versus data
can also be interpreted in the context of the same SMEFT operators~\eqref{eqn:OpsL} 
involving the second and third generation and related simplified models~\cite{Capdevila:2023yhq}.
Together, this provides a strong motivation for many previously studied frameworks to be revisited.
A comprehensive summary of possibly flavor-compatible UV models in this regard exceeds the scope of this review. 
UV paradigms that generate LFUV and flavor-compatible phenomenology include:
\begin{enumerate*}[label = \bf(\roman*)]
	\item minimal flavor violation (MFV);
	\item Froggatt--Nielsen~(FN) frameworks and discrete flavor symmetries; 
	\item dynamical alignment mechanisms, such as rank-one flavor violation or flavor locking;
	\item warped extra dimensions and grand unified theory (GUT) frameworks.
\end{enumerate*} 
We will briefly discuss these below,
in service of highlighting areas or ideas whose further development or revisitation, in our view, would be well motivated.

MFV provides a parsimonious approach to implementing flavor structure 
that can be compatible with data for certain simplified models (see e.g.~\cite{Davidson:2010uu, Fajfer:2012jt, Freytsis:2015qca}).
In this framework, all quark flavor violation is generated by bifundamental spurions 
charged under $U(3)_Q \times U(3)_U \times U(3)_D$ quark flavor symmetries,
i.e. by $Y_u \sim (3, \bar{3}, 1)$ and $Y_d \sim (3, 1, \bar{3})$.
One may instead consider $SU(2)_Y \times SU(2)_U \times SU(2)_D$ subgroup acting on the first two generations only,
with spurions $\Delta_u \sim (2, 2, 1)$ and $\Delta_d \sim (2, 1, 2)$, 
plus an additional spurion $V \sim (2, 1,1)$ to mediate coupling to the third quark generation~\cite{Barbieri:2011ci,Barbieri:2012uh,Barbieri:2014tja}; 
see also Ref.~\cite{Fuentes-Martin:2019mun}.

As an example, one can read off from Fig.~\ref{fig:smpmodels} 
that taking an $S_3$ leptoquark transforming as $\sim (1,1,\bar{3})$ under quark flavor 
produces the vertex $ Q_L^{a \alpha} (Y_d^\dagger)_{\alpha}^{\hat{\alpha}} L_L^{b} \tau^k_{ab} (S^{k}_3)_{\hat\alpha}$. 
In the quark mass basis after EW symmetry breaking, 
this generates the semileptonic decay operators
$\text{Tr}(S_3^3 M_d V_{\text{CKM}}^\dagger u_L) e_L$ and $\text{Tr}(S_3^3 M_d  d_L) \nu_L$ (plus others). 
Here $M_d$ is the diagonal down quark mass matrix, that suppresses couplings to the first generation,
$V_{\text{CKM}}$ is the CKM matrix, 
and the trace is over flavor indices.
A similar game can be applied to the lepton sector.
The flavor-diagonal $d_L \nu_L$ coupling suppresses $\bsnunu$,
but tree-level $\bsnunu$ decays arise once higher-order MFV terms are included, 
leading to potentially tight constraints. 
Many other constraints also apply, including $S_3$ decays to quark-lepton final states,
as well as one-loop corrections to $Z$ decays.
Studies of MFV models that can account for the LFUV anomalies have been examined in e.g. Refs.~\cite{Freytsis:2015qca,Gripaios:2015gra,Aloni:2017ixa},
though mostly including the neutral current LFUV or muon $g-2$ anomalies as additional constraints.
A comprehensive analysis of viable MFV models that are constructed to explain only charged-current LFUV, 
while taking into account most recent precision data, would therefore be welcome.

The principal idea of the FN mechanism~\cite{Froggatt:1978nt} is an additional $U(1)_H$ ``horizontal'' symmetry 
under which the SM fermions have non-universal charge assignments.
An SM-singlet spurion breaking $U(1)_H$ allows the generation of SM-like Yukawa interactions,
with exponential hierarchies in their texture.
The original motivation was to explain the SM mass and mixing hierarchies in the quark sector, 
with a plethora of implementations 
(see e.g. the recent Refs~\cite{Smolkovic:2019jow, Cornella:2023zme} among very many).
It can be extended to explain hierarchies in NP leptoquark models compatible with LFUV anomalies~\cite{Hiller:2016kry,Bordone:2019uzc}
via discrete flavor symmetries~\cite{Hiller:2016kry} (see also e.g.~\cite{Bigaran:2022kkv} that does not require FN),
or by including additional internal horizontal symmetries, spurions, and heavy mediators to further modify the FN hierarchies~\cite{Asadi:2023ucx}.
The versatility of this general framework makes it feasible to account for charged current LFUV alone, without implicating other anomalies,
although sometimes tensions can still arise if the size of the LFUV is too large~\cite{Hiller:2016kry}.

In contrast to MFV, one may posit that the required flavor violation in the NP couplings arises via an outer product of a vector of NP couplings,
generating the so-called rank-one flavor violation hypothesis~\cite{Gherardi:2019zil}. 
For instance, one could imagine $\Wc{VL}^{ab} \sim n^a n^b$, 
where $n^a$ itself is generated dynamically by integrating out heavy NP states in an operator of the class $n_a \Qbar_L^a \Op{NP}$,
with the goal that $n$ is aligned mostly with the third SM generation in the SM quark mass basis.
This approach has been mainly studied in the context of the neutral current $R_{K^{(*)}}$ anomalies or $B \to K \nu\nu$~\cite{Marzocca:2024hua}, however.
Relatedly, the flavor-locking mechanism~\cite{Knapen:2015hia} can achieve the required flavor structure 
by exploiting algebraic properties of renormalizable theories of spurions.
Generic choices of the couplings in the spurionic potential forces its vacuum into aligned, rank-one, and disjoint configurations amenable to realistic theories of flavor~\cite{Altmannshofer:2017uvs},
without needing horizontal symmetries on the SM quarks themselves.
For a $W_1^\prime$ simplified model with a right-handed neutrinos~\cite{Asadi:2018wea, Greljo:2018ogz},
a combination of bifundamental and fundamental spurions under $U(3)_Q \times U(3)_{U,D}$ and $U(3)_{U,D}$ feature a specialized vacuum configuration, 
such that only $\cbar_R \slashed{W}_1^\prime b_R$ operators are generated~\cite{Robinson:2018gza}.
Extending this to other simplified models has not yet been attempted.

Warped extra dimension approaches provide a natural framework for UV completions compatible with the LFUV anomalies~\cite{Megias:2017vdg},
in particular for those involving right-handed neutrinos~\cite{Carena:2018cow},
while also addressing the EW hierarchy problem.
The desired flavor structure arises from localization of fermion generations along the extra dimension itself
(analogous to differing compositeness of the quark (and lepton) species in the dual holographic theory).
This framework generates couplings to Kaluza-Klein modes of the form $V_{u}^\dagger G V_d$, 
in which $V_u^\dagger V_d$ generates $V_{\textrm{CKM}}$ and $G$ is diagonal.
A predictive implementation requires some flavor model of $V_{u,d}$ themselves,
and moreover one must require the lepton mixing to be specially chosen to avoid charged lepton flavor violation.
It is not known whether such a scenario can naturally arise, and solely for charged current LFUV.

Simplified models involving leptoquarks are highly suggestive of grand-unified theory (GUT) UV completions,
in particular the Pati–Salam gauge group, with an associated extensive literature.
Because the GUT scale---the nominal mass scale of the leptoquarks---is required to be $\sim$\,TeV 
account for LFUV anomalies, a source of flavor violation is required to prevent tight flavor constraints from the first SM generation.
For vector leptoquarks, the flavor violation can be achieved  by introducing mixing with vector-like fermions (e.g. \cite{Calibbi:2017qbu}), 
extra-dimension embeddings (e.g.~\cite{Blanke:2018sro}), among many others.
For scalar leptoquarks, by contrast, the GUT scale can be much higher than $M$ while also allowing flavor-violating leptoquark couplings (see e.g.~\cite{Heeck:2018ntp}).
The viability of the many GUT approaches in the literature to generate only charged-current LFUV requires study.

\section{Predictions and phenomenology}
The ultimate phenomenological goal is to perform NP inference on $\bctn$ data: that is, inferring CLs on the space of NP WCs.
This requires a precision description and predictions for the distributions of observables in $\bctn$ and $\bcln$ decays, as well as backgrounds.
In $\bcln$ processes ($l = e$, $\mu$, $\tau$), at leading EW order the amplitudes factorize as
\begin{equation}
	\label{eqn:genampl}
	\mathcal{M} [\Hb \to \Hc l \nu] \propto \MatrixElem{ \Hc}{ \cbar\, \Gamma b }{ \Hb } \, \bar{u}_l \Gamma' v_\nu\,.
\end{equation}
Here $\Hb$ and $\Hc$ denote some beauty and charmed hadron or hadronic system,
and $\cbar\, \Gamma b$ and $\bar{u}_l \Gamma' v_\nu$ are the quark and lepton currents of interest, respectively, arising from Eq.~\eqref{eqn:opsL}.
In Eq.~\eqref{eqn:genampl} we have dropped overall normalizations and WCs.
The leptonic matrix element is written in terms of its tree-level spinor representation, 
with the implicit understanding that real and virtual radiative corrections arise at higher order in QED~\cite{Cima:2025trg}.
The same applies to $\Hb \to \Hu l \nu$.
For all current data, $\Hb$ is either a reconstructed $B$, $B_s$ or $B_c$ meson, or a $\Lambda_b$ baryon.
In exclusive processes, the outgoing hadronic state is fully reconstructed,
while inclusive processes involve final-state charmed or light hadronic systems that are not.

The QCD matrix elements  $\matrixelem{ \Hc}{ \cbar\, \Gamma b }{ \Hb }$ require a theory description in order to obtain predictions for the decay amplitudes and rates.
Because QCD matrix elements cannot be computed from first principles,
these descriptions inherently introduce various types and degrees of systematic hadronic uncertainty.
In the SM, the difference between $\Hb \to \Hc \tau \nu$ and $\Hb \to \Hc \ell \nu$ differential or total rates arise at $\mathcal{O}(m_{\tau}^2/q^2)$,
and thus the leading hadronic uncertainties cancel in ratios of semitauonic versus semileptonic processes,
as does the CKM matrix element $|V_{cb}|$ (or $|V_{ub}|$).
For  LFUV, ratios such as  
\begin{equation}
  R(\Hc) = \BR{\Hc \to \Hc \tau \nu}/ \BR{\Hb \to \Hc \ell\nu}\,,\qquad \ell = e, \mu \, ,
\end{equation}
thus feature comparatively clean and more precise theoretical SM predictions compared to the absolute differential rates.
E.g. at present the precision of $R(\Dx)$ predictions from fits using LQCD and experimental inputs is $\sim 1\%$ (LQCD alone is $\sim 3\%$) 
versus $\sim 4\%$ for the $\Bbar \to \Dx \ell \nu$ branching fraction normalizations.

Of crucial importance is that theory-based hadronic uncertainties also affect the experimental measurements themselves.
This is because in practice one must compare yields with a signal plus background fit that is sensitive to the theory-derived signal 
and background templates and their underlying hadronic uncertainties 
(and also the NP model itself; see Sec.~\ref{sec:worldav}).
Put in other words, even if the leading hadronic uncertainties cancel in the theory predictions for $R(\Hc)$, 
they may not necessarily do so in the experimental measurement of these quantities.
Thus NP inference from $\bctn$ data may be 
ultimately systematically theory-limited by the precision of hadronic descriptions of the SM and NP differential rates.

In this section we review the fundamentals of theory descriptions for exclusive and inclusive processes
that are relevant to NP inference, and main sources of their various limitations, caveats, and uncertainties.
Inclusive final states are denoted hereafter as $X_{c,u}$, and $X = X_c + X_u$.
Beyond the scope of this review are semi-inclusive processes, that have $\Hc X_u$ final states and may also be candidate NP signal modes,
as well as doubly-inclusive data from future tera-$Z$ colliders, in which the initial beauty state is itself not fully reconstructed.
See the very recent Ref.~\cite{DeFazio:2026gju} for examples of more technical discussions, not possible in a review of this length, on the state of the art for some topics covered below.
We emphasize our focus is on NP inference, and thus omit much discussion of SM-based state-of-the-art theory treatments.

\subsection{Exclusive modes and form factors}

The hadronic matrix element can always be decomposed schematically as
\begin{equation}
	\label{eqn:FFdec}
	\MatrixElem{ H_{c} }{ \cbar\, \Gamma b }{ \Hb } = \!\! \sum_{J^P\!(L)} ~ \sum_{\alpha \in J^P\!(L)} \!\!\! F_\Gamma^{\alpha}(q^2) \, \MatrixElem{ H_{c} }{ \cbar\, \Gamma b }{ \Hb }_{\alpha}\,,
\end{equation}
in which each matrix element in the sum is a partial wave amplitude (PWA): 
a matrix element in which the external hadronic state has definite total and orbital angular momentum and parity $J^P(L)$.
There may be several independent PWAs for each $J^P(L)$ configuration, indexed by $\alpha$ in Eq.~\eqref{eqn:FFdec}.
Their counting follows from selection rules~\cite{Ji:2000id, Papucci:2021pmj}.
Each PWA can be expressed in terms of spacetime representations of the external states---i.e. 
involving momentum, polarizations, spinors, tensors etc as appropriate---and
in turn is associated with a unique form factor (FF), $F_\Gamma^{\alpha}$. 
An FF is itself a (Lorentz-invariant) QCD amplitude.
For single-hadron transitions, the FFs depend only on the recoil $q^2 = (p_{\Hb} - p_{\Hc})^2$ (and implicitly also $m_{\Hb}$ and $m_{\Hc}$).
The $PT$ symmetry of QCD ensures that the FFs are real
and required $PT$ and conjugation transformation relations fix relative factors of $i$ in the spacetime representations of the PWAs~(see e.g. Ref.~\cite{Manzari:2024nxr}).
(Some literature multiplies the FF definitions by an overall factor of $\pm i$ or $\pm1$.)

Depending on the underlying theory framework used (see below) other bases of spacetime representations 
and thus FFs may be more convenient versus a PWA decomposition.
As the most precise data are/will be from $\Bbar \to \Dstar$ decays, 
we provide here as an example three common FF basis choices for $\Bbar \to \Dstar$ for both the SM (cf.~\cite{Bernlochner:2021vlv}) and NP,
respectively the ``spectroscopic'', heavy quark (HQ) symmetry, and helicity bases:
\begin{subequations}
\label{eqn:BDsFFs}
\begin{align}
\MatrixElem{D^*}{\cbar \g^5 b}{\Bbar} 
	& = a_0 (\epsilon^* \!\cdot\!  p) =  -h_P(\epsilon^* \!\cdot\!  v)  =  A_P (\epsilon^* \!\cdot\!  q)\,, \\
\MatrixElem{D^*}{\cbar \g^\mu b}{\Bbar} 
	& = -2 i \widetilde{g}\, \varepsilon^{\mu\nu\alpha\beta}\,\epsilon^*_{\nu}p'_\alpha p_\beta 
	 = -i\sqrt{m m'}\, h_V\, \varepsilon^{\mu\nu\alpha\beta}\,\epsilon^*_{\nu}v'_\alpha v_\beta \\
	& = -2 i V (m + m')^{-1}\varepsilon^{\mu\nu\alpha\beta}\,\epsilon^*_{\nu}p'_\alpha p_\beta \nn \\
\MatrixElem{D^*}{\cbar \g^\mu \g^5 b}{\Bbar}  
	& = f \epsilon^{*\mu} + a_+ \epsilon^* \!\cdot\! p\, (p + p')^\mu+ a_- (\epsilon^* \!\cdot\!  p) q^\mu  \\
	& = \sqrt{m m'}\, \big[h_{A_1} (w+1)\epsilon^{*\mu}  - h_{A_2}(\epsilon^* \!\cdot\!  v)v^\mu  - h_{A_3}(\epsilon^* \!\cdot\! v)v'^\mu \big]\,, \nn \\
	& = A_1(m + m')\epsilon^{*\mu} -A_2  \frac{\epsilon^* \!\cdot\! p(p + p')^\mu}{m + m'} + 2m' q^\mu(A_0 - A_3)( \epsilon^* \!\cdot\! p)/q^2 \,, \nn \\
\MatrixElem{D^*}{\cbar \sigma^{\mu\nu} b}{\Bbar}  
	& = -\varepsilon^{\mu\nu \rho \sigma} \bigg[ a_{T_+} \epsilon^*_\rho (p + p')_\sigma + a_{T_-} \epsilon^*_\rho\, q_\sigma + a_{T_0} (\epsilon^* \!\cdot\!  p) (p + p')_\rho\, q_\sigma \bigg]\\
	& = \sqrt{m m'} \varepsilon^{\mu\nu\alpha\beta} \big[ h_{T_1} \epsilon^*_{\alpha}(v+v')_\beta + h_{T_2} \epsilon^*_{\alpha} (v-v')_\beta + h_{T_3}(\epsilon^*\cdot v) v_\alpha v'_\beta\big] \nn \\
	&  = \varepsilon^{\mu\nu \rho \sigma} \Big[ T_1 \epsilon^*_\rho (p + p')_\sigma - (T_1 - T_2)\Big[\big(m^2 - {m'}^2\big) / q^2\Big]\epsilon^*_\rho q_\sigma \nn \\
	&  \quad + (p + p')_\rho q_\sigma \big[(\epsilon^* \!\cdot\!  q)/ q^2 \big] \Big((T_1 - T_2)  - T_3 \big[ q^2 / \big(m^2 - {m'}^2\big)\big]\Big)\Big]\,, \nn
\end{align}
\end{subequations}
and $\matrixelem{D^*}{\cbar  b}{\Bbar} = 0$, by angular momentum and parity conservation.
Here $p^{(\prime)}$, $m^{(\prime)}$, and $v^{(\prime)} = p^{(\prime)}/m^{(\prime)}$ are the momentum, mass, and velocity of the $\Bbar$ $(\Dstar)$,
while $\epsilon$ is the $\Dstar$ polarization, the recoil $q = p - p'$ 
and the recoil parameter $w = v \cdot v' = (m^2 + {m'}^2 -q^2)/(2 m m')$,
which is also the $\Dstar$ boost in the $\Bbar$ frame.

In the helicity basis, there are two derived FFs 
$A_{12}  = [A_1(m \!-\! {m'}^2 \!\!-\!q^2)(m \!+\! m')^2 \!-\! 4A_2|p'|^2 m^2]/[16m {m'}^2(m + m')]$ and 
$ T_{23}  = [T_2(m^2 \!+\! 3{m'}^2 \!\!-\!q^2)(m^2 \!-\! {m'}^2) \!-\! 4T_3|p'|^2 m^2 ]/[8 m {m'}^2(m - m')]$,
where the $\Dstar$ momentum $|p'| = m'\sqrt{w^2-1}$. 
Much literature uses the SM FFs $P_1$ and $F_1$, related to the $A_i$ via 
$P_1 =  2 A_0 \sqrt{m m'}/(m + m')$ and $F_1 = 8 m m' A_{12}$.
These FFs arise in the helicity amplitudes, and are used mainly in dispersive methods-based frameworks (see below).
(Note also $T_1(q^2 = 0) = T_2(0)$ and $A_0(0) = A_3(0)$,
in order to regulate spurious $1/q^2$ divergences.)

There are subtle but important sign conventions and transformations in FF definitions
that can be a source of confusion and must be treated carefully.
The axial-tensor matrix element, $\matrixelem{\Hc}{\cbar \sigma^{\mu\nu} \g^5 b}{\Hb}$, 
is related to the tensor matrix element by the duality identity 
\begin{equation}
	\label{eqn:sigdual}
	\sigma^{\mu \nu} \g^5 \equiv \pm (i/2) \varepsilon^{\mu \nu \alpha \beta} \sigma_{\alpha \beta}\,,
\end{equation}
whose sign choice is equivalent to that in $\text{Tr}[\g^\mu \g^\nu \g^\alpha \g^\beta \g^5] \equiv \mp 4i \varepsilon^{\mu \nu \alpha \beta}$.
For the now-standard definition $\g^5 \equiv +i \g^0 \g^1 \g^2 \g^3$, this is in turn equivalent to the choice $\varepsilon^{0123} = \pm 1$.~\footnote{%
Overall there are four possible sign convention choices---the signs of $\g^5$ and $\varepsilon^{0123}$---but 
what matters is their relative sign, as one can always uniformly rephase all the matrix elements.
Thus, by always applying the standard $\g^5$ definition, one can consider just the sign of the Levi-Civita symbol without loss of generality.
This sign manifests also in the sign of the $\varepsilon^{\mu\nu\rho\sigma} \g^5$ term in the Chisholm identity. 
Note further $\varepsilon^{0123} = - \varepsilon_{0123}$, which can sow even more confusion.}
Under this rubric, most literature chooses $\varepsilon^{0123} = +1$, which is also the convention chosen in Eq.~\eqref{eqn:BDsFFs}. 
However, the influential Ref.~\cite{Manohar:2000dt} chose the opposite sign for its $\Bbar \to \Dstar$ FFs, 
which was also followed by e.g. Refs~\cite{Bernlochner:2017jka, Bernlochner:2021vlv}.
The overall effect in practice for $\Bbar \to \Dstar$ is to flip the sign of the definition of the vector and tensor currents, which involve the Levi-Civita symbol, 
with respect to the pseudoscalar, axial-vector and axial-tensor currents, which remain the same in either choice.
For other $\Hb \to \Hc$ processes, one should beware of sign discrepancies in FF definitions via the Levi-Civita symbol sign choice.

Some literature instead defines the tensor FFs via the matrix elements for $\cbar \sigma^{\mu\nu} q_\nu b$ and $\cbar \sigma^{\mu\nu} \g^5 q_\nu b$ together.
The duality identity~\eqref{eqn:sigdual} plus self-dual properties of the tetrad formed by $q^\mu$ and its three transverse polarizations
ensure these definitions are equivalent to the above~\cite{Papucci:2021pmj}.~\footnote{%
This definitional scheme introduces explicit hadron mass dependence into HQET matrix elements and violates reparametrization invariance,
mandating careful bookkeeping to match the FFs onto HQET beyond leading order.}
Furthermore, $C$ conjugation changes the sign of the vector, tensor and axial-tensor currents.
The FF definitions for the conjugate $\Hbbar \to \Hcbar$ process thus involve a sign flip in the corresponding matrix elements 
versus those for $\Hb \to \Hc$ (see e.g. \cite{Ligeti:2016npd} for $\Bbar \to \Dx$).
A further subtlety can arise once the PWA spin becomes sufficiently high:
in the NP tensor current in $\Lambda_b \to \Lambda_c^*(3/2^-)$, the PWA decomposition becomes overcomplete, 
in the sense that one linear combination of FFs does not enter the physical amplitudes. 
This has important consequences for matching of FFs onto HQET in particular~\cite{Papucci:2021pmj}.

In the following list, organized by  $J^{P}\big|_{\text{in}} \to J^{P}\big|_{\text{out}}$ of the external hadrons, 
we provide a directory of references which define FF basis conventions for all NP currents 
for a variety of exclusive $\Hb \to H_{c,u}$ semileptonic decays that may probe NP with current or future data.
References are grouped by spectroscopic, HQ symmetry and helicity bases, respectively.
\begin{itemize}[labelwidth=20pt, labelindent=10pt, noitemsep, topsep =5pt, leftmargin=78pt, font=\bfseries]
	\newcolumntype{L}{ >{\raggedright\arraybackslash $} m{4cm} <{$}}
	\newcolumntype{M}{ >{\raggedright\arraybackslash $} m{2.5cm} <{$}}
	\newcolumntype{C}{ >{\centering\arraybackslash} m{1.5cm} <{}}
	\item[$0^- \to 0^-$:] 
			\begin{tabular}{LCCC} \Bbar \to D, \pi;\ B_c^- \to \eta_c, \Dbar  				&	\cite{Ligeti:2016npd,Gubernari:2018wyi} 	& \cite{Bernlochner:2017jka}	& --- \\
			\end{tabular} \\[-3pt]
	\item[$0^- \to 1^-$:] 	
			\begin{tabular}{LCCC} \Bbar_{(s)}  \!\to\! \Dstar_{(s)}, \rho, \omega;\ B_c^- \!\to\! J/\psi	& 	\cite{Ligeti:2016npd}				& \cite{Bernlochner:2017jka}  	& \cite{Bharucha:2015bzk, Gubernari:2018wyi} \\
			\end{tabular}  \\[-3pt]	
	\item[$0^- \!\to\! (0,1,2)^+$:]		
			\begin{tabular}{LCCC} \Bbar \to D^{**} 				& --- 												& \cite{Bernlochner:2017jxt} 				& \cite{Cheng:2004yj} \\
			 \end{tabular} \\[-3pt]
	\item[$\frac{1}{2}^+ \to \frac{1}{2}^+$:]
			\begin{tabular}{LCCC} \Lambda_b \to \Lambda_c, p 		&  \cite{Chen:2001zc}				& \cite{Bernlochner:2018bfn} 	& \cite{Feldmann:2011xf, Detmold:2016pkz} \\	 
			\end{tabular} \\[-3pt]
	\item[$\frac{1}{2}^+ \to (\frac{1}{2}, \frac{3}{2})^-$:]	
			\begin{tabular}{LCCC} \Lambda_b \to \Lambda_c^{*}		& ---			& \cite{Papucci:2021pmj} 		& \cite{Descotes-Genon:2019dbw, Meinel:2020owd, Meinel:2021rbm} \\
			\end{tabular}						
\end{itemize}
In some cases, basis conventions were first introduced for other exclusive hadron transitions, 
such as in $b \to s$ decays, but involving the same $J^{P}\big|_{\text{in}} \to J^{P}\big|_{\text{out}}$ of the external hadrons as those considered here.
State-of-the-art predictions, whose coverage is outside our scope here, typically cite at least one of these references.

Finally, in $\bctn$ or $\butn$ decays, the $\tau$ itself may decay hadronically to the first two SM generations,
and the corresponding matrix elements require an FF representation.
It is generically assumed---and we shall assume here---that such processes proceed through the SM only. 
See e.g. Ref.~\cite{Shekhovtsova:2012ra} for a treatment of hadronic $\tau$ decays.

\subsection{Theory frameworks and parametrizations}
\label{sec:thframe}
The FFs cannot be computed in the continuum from first principles in QCD itself.
However, approaches based on
\begin{enumerate*}[label = \bf(\roman*)]
	\item LQCD,
	\item HQET,
	\item dispersive methods,
	\item quark models,
	\item and/or QCD sum rules (QCDSR) and light-cone sum rules (LCSR)
\end{enumerate*}
can be used to generate SM and NP predictions. 
Typically, one considers a parametrization of the FF functions---a ``FF parametrization''---based on one or more of these approaches and seeks predictions for its parameters.
A full exposition of these methods requires its own review; 
see e.g. Ref.~\cite{Khodjamirian:2020btr} and references therein.
Here we provide some brief commentary relevant to NP inference both now and in a future precision era.
Table~\ref{tab:THframes} summarizes the main advantages and drawbacks of these theoretical frameworks.

\begin{table}[t]
\let\tabvshift\relax
%Vertically space the first, last columns with middle top aligned. Argument is max number of rows in itemize cells
\newcommand{\settabvshift}[1]{
	\gdef\tabvshift{\vspace{\dimexpr 0.5\baselineskip * (#1 - 2)\relax}}
}
\newcolumntype{L}{ >{\raggedright\arraybackslash} p{4.5cm} <{}}
\newcolumntype{M}{ >{\raggedright\arraybackslash} p{3.25cm} <{}}
\newcolumntype{C}{ >{\tabvshift\centering\arraybackslash} p{1cm} <{}}
\newcolumntype{D}{ >{\tabvshift\raggedright\arraybackslash} p{2cm} <{}}
\newenvironment{tableitemize}{
	\vspace{-3pt} 
	\begin{itemize}[wide, leftmargin=*, labelwidth=!, labelindent=-10pt, noitemsep, topsep =0pt] 
	}{%
	\vspace{3pt}
	\end{itemize}
	\ignorespacesafterend
	}
\renewcommand{\arraystretch}{1}
\begin{threeparttable}
\scalebox{0.73}{\parbox{1.4\linewidth}{
\begin{center}
\begin{tabular}{DMLLC}
	\hline
	Framework & Concept & Features & Limitations/Challenges & NP \settabvshift{6}\\[3pt]
	\hline\hline
	\textbf{LQCD} 	& \begin{tableitemize} 	\item QCD on finite and discretized Euclidean spacetime
									\end{tableitemize} 	
				& \begin{tableitemize} 	\item Compute SM and NP matrix elements
									\item Precision predictions at zero recoil ($w=1$)
									\end{tableitemize}  
				& \begin{tableitemize} 	\item Systematics: continuum extrapolation, finite volume, quark mass effects, \ldots
							  		\item QED, isospin corrections $\sim 1\%$
									\item Current tensions vs $\Bbar \!\to\! D^*\ell\nu$ data at $w>1$
									\end{tableitemize} 
				&   Yes \settabvshift{4} \\  
				\hline
	\textbf{HQET} 	& \begin{tableitemize} 	\item EFT of QCD for $m_Q \gg \Lqcd$
									\item Power expansion in $1/m_Q$, $\alpha_s$.
									\end{tableitemize} 
				& \begin{tableitemize} 	\item Relations between SM/NP
									\item $\ell$ vs $\tau$ relations
									\item Zero-recoil constraints, HQ supermultiplet relations
									\end{tableitemize}
				& \begin{tableitemize} 	\item Proliferation of second-order power corrections $\sim 4\%$
									\item $\alpha_s^2$ contributions
									\item Renormalon ambiguities
									\end{tableitemize}				 
				& Yes \settabvshift{3} \\ 
				\hline
	\textbf{Dispersive} 
				&  \begin{tableitemize} 	\item Unitarity and analyticity
									\item QHD
									\end{tableitemize}
				& \begin{tableitemize} 	\item Unitarity bounds
									\item Conformal $z$ expansion 
									\item Constrained FF forms 	
									\end{tableitemize}					
				& \begin{tableitemize} 	\item No relations between SM/NP
									\item No $\ell$ vs $\tau$ relations
									\item Subthreshold branch points
									\end{tableitemize}
				& No\tnote{a} \settabvshift{2} \\
				\hline 
	\textbf{Quark models} 
				& \begin{tableitemize} 	\item Variety of constituent quark frameworks
									\end{tableitemize}
				& \begin{tableitemize}	\item Predictive, constrained, and/or parameter free
									\end{tableitemize}
				& \begin{tableitemize}	\item	 Hadronic model dependent
									\item Uncertainties unknown
									\end{tableitemize}
				& Yes\tnote{b} \settabvshift{6} \\ 
				\hline 						
	\textbf{Sum rules} 
				& \begin{tableitemize} 	\item QCDSR: analytic properties of 3-pt correlators plus OPE
									\item LCSR: 2-pt correlator with OPE near the light-cone
									\end{tableitemize}
				& \begin{tableitemize} 	\item QCDSR: Zero-recoil predictions for IW functions
									\item LCSR: twist-ordered expansion. Predictions for $b \to u$ ($b\to c$) at high recoil (beyond kinematic endpoint)
									\end{tableitemize}
				& \begin{tableitemize}	\item Systematics: QHD violation, Borel parameters, scale variations,  $q^2$ correlated
									\item LCSR: non-pert modeling, twist truncation, extrapolation from unphysical recoil.
									
									\end{tableitemize}
				& Yes\tnote{c} \\
				\hline			
\end{tabular}
\end{center}
\begin{tablenotes}
      \item[a] Used to parametrize LQCD NP predictions, via fits to synthetic data
      \item[b] Typically predictions do not have quoted uncertainties
      \item[c] QCDSR predictions exist for IW functions, applicable to NP predictions via HQET
\end{tablenotes}
}}
\vspace{3pt}
\caption{Summary of main concepts, features, and limitations of the theoretical frameworks used to describe or predict exclusive SM and NP  form factors.}
\label{tab:THframes}
\end{threeparttable}
\end{table}

\textbf{Lattice QCD:} LQCD directly computes hadronic matrix elements by implementing QCD on a discretized Euclidean spacetime.
This first-principles approach may eventually permit subpercent precision NP predictions once QED and isospin effects are incorporated.
There are, however, potentially large systematic uncertainties from the required chiral continuum extrapolation, 
finite-volume and quark mass effects,
as well as excited state contributions, that are hard to assess. 
At present,  the LQCD results are most precise at lower recoil:
Predictions at zero recoil ($w=1$) in $\Bbar \to D^*$ and $w\ge 1$ for $\Bbar \to D$ are considered particularly reliable, 
and are critical, widely used inputs to FF fits.
At $w>1$ for $\Bbar \to \Dstar$ there is tension with experimental data (see Sec.~\ref{sec:np-vs-hadronic})
and the precision of LQCD predictions is currently somewhat lower than from fits (based on other theoretical approaches) to experimental data.
FNAL/MILC~\cite{FermilabLattice:2021cdg}, HPQCD~\cite{Harrison:2023dzh} and JLQCD~\cite{Aoki:2023qpa} all provide SM FFs beyond zero recoil. 
The FNAL/MILC and JLQCD simulations extend over about a quarter of the physical recoil range from $w=1$, 
while the HPQCD results covers the full $q^2$ range and also includes all NP FFs.
At present, e.g., the precision of the normalization of the $\Bbar \to \Dstar$ tensor FFs is $\simeq 2\%$,
while higher-order parameters---characterizing higher-order derivatives in recoil---carry much larger uncertainties of $20$-$90\%$.
For doubly heavy hadron transitions, such as $B_c \to \eta_c$ or $J/\psi$, 
LQCD provides the sole source of precise NP predictions~\cite{Harrison:2025yan}.

\textbf{HQET parametrizations:} 
At present continuum methods provide the most precise predictions for NP matrix elements:
even in a far futuristic era of subpercent precision LQCD predictions for e.g. $\Bbar \to \Dx$, 
they will continue to provide comparable predictive power or crosschecks, 
and leading predictive power in the case of resonant or excited state processes.
HQET~\cite{Eichten:1989zv,Georgi:1990um} is an EFT derived from QCD in the regime $m_Q \gg \Lqcd$. 
A mass-subtracted field redefinition plus HQ velocity projection 
allows for a reorganization of QCD into a power expansion in $1/m_Q$ and $\alpha_s$,
in which the leading terms preserve HQ spin and flavor symmetries.
QCD matrix elements are matched onto HQET matrix elements order-by-order.
The form factors of HQET matrix elements are Isgur-Wise (IW) functions. 
HQ symmetry embeds certain pairs of hadrons into HQ supermultiplets that represent the external states,
and at lower orders there are fewer IW functions than QCD FFs.
This allows for powerful relations not only between all SM and NP matrix elements, and $\tau$ and $\ell$ modes, but also between supermultiplet pairs such as  $\Bbar \to \Dstar$ and $\Bbar \to D$.

The paradigmatic and systematically improvable approach to obtain NP predictions for the $\Hb \to \Hc \tau \nu$ modes is to fit 
an HQET-based parametrization to precision measurements of the light leptonic $\Hb \to \Hc \ell \nu$ decays, 
that are assumed to be SM only, to recover the IW parameters.
LQCD inputs are used to further constrain fits, mostly at low recoil.
This approach works e.g. for $\Bbar \to \Dx$ at first order in HQET, and to second order for $\Lambda_b \to \Lambda_c$.
For $\Bbar \to \Dx$,  the ``BLPR'' parametrization~\cite{Bernlochner:2017jxt} generates NP predictions in this manner.
 In $\Bbar \to \Dx \tau \nu$ at second order, the number of IW functions greatly exceeds the number of FFs, 
 leading to unknown $\sim 4\%$ hadronic uncertainties
that cannot be constrained from $\Bbar \to \Dx\ell\nu$ data. 
New methods or ansatzes, however, appear to permit control of these second-order corrections~\cite{Bernlochner:2022ywh},
but require establishment from first principles.
Additional challenges include control of $\alpha_s^2$ contributions, 
and cancellation of higher-order renormalon ambiguities within the chosen HQ short distance scheme.
Further, incorporation of known chiral corrections is required to account for leading $U$-spin flavor-breaking effects,
which would in turn allow for combined treatments of $B \to \Dx l \nu$ and $B_s \to \Dx_s l\nu$ data.

\textbf{Dispersive parametrizations:} Dispersive methods exploit analyticity and unitarity of the hadronic matrix elements to obtain powerful constraints on their structure.
For a definite-$J^P$ current, a (multiply)~subtracted dispersion relation in $q^2$ can be constructed involving the imaginary part of its two-point correlator.
Matching partonic and hadronic representations of this dispersion relation via quark-hadron duality (QHD) allows one to obtain a dispersive bound,
that can be expressed in terms of weight functions times squared FFs in a PWA decomposition, i.e. the helicity basis.
Analytic properties of a conformal mapping $q^2 \mapsto z \in \mathbb{C}$ 
allow each helicity form factor to be expressed as an analytic function in $z$ divided by a calculable, FF-specific weight function and $J^P$-specific Blaschke factors.

This approach provides a constrained functional form for the FFs combined with a naturally truncatable expansion in $z$, 
as $|z| \ll 1$ in the kinematically allowed recoil region.
Furthermore, the coefficients of the analytic function must satisfy a unitarity bound, at least on a statistical level.
FF parametrizations constructed this way are often referred to as ``BGL''~\cite{Boyd:1997kz}.
There is no intrinsic relation between matrix elements for currents with different $J^P$, nor between SM and NP.
Thus one cannot generate NP predictions from SM fits to the light hadrons, nor even predictions for $\Bbar \to \Dstar \tau \nu$ in the SM,
as the latter involve the $P_1$ form factor that cannot be constrained from $\Bbar \to \Dstar \ell \nu$ data. 
For excited state or $B_{c,s}$ decays, 
the decay's kinematic threshold lies well above the hadronic pair production threshold for each definite-$J^P$ current leading to subthreshold branch points.
These distort the conformal mapping: treatments of the derived unitarity bounds differ~\cite{Gubernari:2023puw,Simula:2025fft}.
Nonetheless, because of their highly generic structure, dispersive BGL-style parametrizations 
are usually chosen for fitting LQCD synthetic data including NP predictions~\cite{Harrison:2023dzh, Harrison:2025yan}.

\textbf{Sum rules:} Additional inputs may come from QCD sum rules (QCDSR) and light-cone sum rules (LCSR). 
Both of these frameworks consider correlators of interpolating currents plus weak currents,
using an operator product expansion (OPE) to express them in terms of perturbatively computable kernels times non-perturbative condensates or distribution amplitudes.
In the case of LCSRs (see \cite{Khodjamirian:2023wol} for a review) the product of currents is expanded near the light cone,
resulting in a twist-ordered expansion.
For $b \to u$ decays, LCSRs are expected to converge quickly at high-recoil.
Both frameworks feature unknown systematics from QHD violation, scale variations, choice of Borel parameters, and continuum thresholds,
and results are correlated across the $q^2$ range.
LCSRs further require model-dependent inputs for non-perturbative elements and control of twist truncation effects.
Long-available QCDSR results for leading and subleading IW functions in the $\Bbar \to \Dx$ system can be used as additional constraints on SM or NP predictions,
i.e. via HQET. 
These results are highly correlated and the uncertainties are large, and thus their role in future precision predictions, whether SM or NP, is doubtful.
LCSRs provide SM and NP predictions for $b \to u$ systems, in which the boost of the light partons is typically large, 
including higher-twist contributions~\cite{Leljak:2021vte, Bharucha:2015bzk}.
SM and NP LCSR predictions are also available for $b \to c$ processes (see Ref.~\cite{Gubernari:2026sqc} and therein).
These must be computed in an unphysical $q^2 < 0$ regime far from the kinematic endpoint, 
in which unknown higher-order (twist and parton multiplicity) contributions are thought to be suppressed.
Extrapolation is then required to the physical $q^2$ regime, which may introduce additional sources of systematic uncertainty and some model dependence.

\textbf{Quark models:} NP predictions may also be derived from constituent quark models, 
that use a variety of hadronic model dependent assumptions and setups.
Examples include the covariant confined quark model, relativistic confined quark model, and the light front quark model,
all of which may be used to generate NP predictions. 
While these models have the advantage of being highly predictive, the inherent systematic uncertainties associated with model dependence are unknown:
often predictions are quoted without uncertainties, and thus at best provide an intuition check of the NP predictions discussed above.

\subsection{Inclusive modes}

In inclusive $\Hb \to X_{c,u} l \nu$ decays, the final hadronic state is ill-defined,
preventing a FF-based approach.
Instead, one relates the fully differential rate to the imaginary part of the $\Hb$ forward-scattering matrix elements per the optical theorem,
computing the latter within an OPE with respect to local HQ operators.
The triply differential rate is often expressed in terms of the coordinate system $\hat{q}^2 = q^2/m_b^2$, $y = 2E^*_l/m_b$ and $x = 2E^*_\nu/m_b$ (or $v \cdot q/m_b$),
where $E^*_{l,\nu}$ are the charged lepton and neutrino energies in the $\Hb$ frame.

The OPE breaks down near maximal $\hat{q}^2$ and $E_l$.
This is of particular importance for $\Bbar \to X_u \ell \nu$ decays, where data must be taken in the high $E_\ell$ regime, 
beyond the kinematic endpoint of the otherwise dominant $\Bbar \to X_c \ell \nu$ background.
The problematic terms can be resummed in terms of non-local light cone operators.
This amounts to a multiscale OPE, leading to factorization of the differential rate into a hard scattering kernel at $\mathcal{O}(m_b^2)$,
a universal jet function at $\mathcal{O}(m_b \Lqcd)$ and non-perturbative shape functions at  $\mathcal{O}(\Lqcd^2)$.
There are various approaches used to compute the components of this factorization:
see Ref.~\cite{Fael:2024rys} for a recent review. 
Systematic uncertainties arise from parametrization of the shape functions, QHD violation effects, higher-order HQ parameters, and $\mathcal{O}(\alpha_s^2)$ corrections.

For $b \to c \tau \nu$ inclusive decays, the finite $\tau$ mass introduces an additional scale, modifying the Dalitz space by bounding $q^2$ from below.
Treatments are available that include this effect, 
yielding SM predictions for the $E_\tau$ ($\hat{q}^2$) distribution at $\mathcal{O}(\alpha_s/m_b^2)$~\cite{Ligeti:2014kia} 
($\mathcal{O}(\alpha_s/m_b^3)$~\cite{Moreno:2022goo}).
Results for the $b \to u \tau \nu$ $q^2$ distribution can be obtained by taking $m_c \to 0$; see \cite{Ligeti:2021six} for the lepton energy distribution in this limit,
which has singularities.
While most inclusive literature focuses on SM predictions, NP predictions~\cite{Goldberger:1999yh}
are now known at $\mathcal{O}(\alpha_s/m_b^3)$~\cite{Fael:2022wfc} in terms of $E_l$ and $q^2$ moments of the differential rate.

The predicted total inclusive branching fraction must bounded from above the sum of predicted exclusive mode branching fractions. 
Conversely, one may approximate the inclusive prediction as a sum over exclusive modes.
Approximately $14$--$20\%$ of the $\bctn$ rate is not saturated by the $D$ and $\Dstar$ final states~\cite{HeavyFlavorAveragingGroupHFLAV:2024ctg},
so that inclusive measurements are inherently sensitive to a wider range of hadronic final states.

\subsection{Optimal observables for NP inference}
\label{sec:optobs}
Beyond the theoretical cleanliness of the (SM) LFUV ratios via the cancellation of leading hadronic uncertainties,
they are moreover experimentally cleaner than differential rates, 
because of the cancellation of some experimental systematic uncertainties in the ratio.
Observables constructed from ratios of yields---also called shape observables---within 
the same mode can similarly yield cleaner measurements, 
that can be compared with theoretical expectations. 
Common examples of cancelling systematic uncertainties include the lepton particle ID or a tagging efficiency.

Along these lines, polarization and other angular observables, based on differential shapes, are of particular experimental interest.
For example, for a subdecay system $\Hc \to D h$, where $h$ is a light hadronic system, 
the marginal differential decay rate for $\Hb \to (\Hc \to Dh) \tau \nu$ with respect to the $\Hc$-frame polar helicity angle $\theta_h$
can be expressed as $(1/\Gamma) d\Gamma/d\cos\theta_h = \sum_\lambda F_{h,\lambda}(\Hc) D_\lambda(\theta_h)$ 
where $\lambda$ is the helicity of $\Hc$, $D_\lambda(\theta_h)$ are Wigner-$d$ functions 
and $F_{h,\lambda}(\Hc)$ are NP model-dependent polarization fractions, which can be measured from this distribution and compared with theory predictions.
The same can be done for $\Hb \to \Hc(\tau \to h \nu)\nu$, in which $(2/\Gamma) d\Gamma/d\cos\theta_h  = 1 + \alpha_h P_\tau(\Hc) \cos\theta_h$,
where $\alpha_h$ is a computable analyzing power and $P_\tau(\Hc)$ a  NP model-dependent polarization.
Some examples and associated SM predictions are reviewed in e.g. Refs~\cite{Asadi:2020fdo, Bernlochner:2021vlv, Klaver:2024xdy, Iguro:2024hyk}: 
for $\Bbar \to \Dstar \tau \nu$ a typical set of observables considered are $R(\Dx)$, the $\Dstar$ longitudinal polarizations $F_{L,l}(\Dstar)$, $l = \tau$ or $\ell$, and $\tau$ polarizations $P_{\tau}(\Dx)$.

Some of these observables exhibit particular sensitivity or insensitivity to certain NP WCs, 
mainly either because of selection rules (in $J^P$ or HQ symmetry)
or lifting of chiral suppression versus the SM.
A widespread practice assesses compatibility of the (exclusive) data with NP predictions via global fits to sets of these shape observables: 
see e.g.~\cite{Blanke:2018yud, Iguro:2024hyk} among very many others;
cf. Ref.~\cite{Carvunis:2025vab} for $\Bbar \to X_c \ell \nu$.
These draw in data not only from $\bctn$ processes, but also $B_c \to \tau \nu$ and $b \to s$ processes that are related within a particular simplified model, 
in order to derive allowed CLs in the corresponding WC spaces.
Other approaches consider the set of angular moments in the full differential distribution for $\Hb \to (\Hc \to \Hc^\prime h^\prime)(\tau \to h \nu) \nu$,
defined with respect to a basis of orthogonal functions in the angular phase space~\cite{Duraisamy:2013pia,Colangelo:2018cnj,Becirevic:2019tpx,Mandal:2019vwq,Asadi:2020fdo}.
Their measurement can be fit against to NP predictions to similarly obtain CLs on WCs of interest.

It is now broadly appreciated, however, that this practice is susceptible to biases,
arising from mismatches between the NP (hadronic) model template used in the experimental MC framework 
versus the NP (hadronic model) used for inference~\cite{Bernlochner:2020tfi} (see Sec.~\ref{sec:worldav}).
(Whether or not NP and hadronic effects themselves factorize may depend on the framework: see Sec.~\ref{sec:np-vs-hadronic}.)
Because the underlying short-distance physics, SM or NP, is measurement-universal, 
a more robust and unbiased approach is to fold the NP and/or hadronic models into the fully differential data, and recover CLs from fits.
An immediate question, that will become more acute with more data and higher precision, 
is how to balance exploitation of the statistical power of the differential data
versus selection of observables (or regions of phase space) that reduce experimental systematics, such as shapes.
This is a profound optimization problem, that is likely also NP and hadronic model dependent.

In an idealized scenario in which the signal and all background contributions, 
including detector resolution and efficiency, are perfectly modeled, 
an extended unbinned maximum-likelihood fit to the full set of reconstructed observables would be asymptotically efficient, 
i.e. it would extract the maximal statistical information on the parameters of interest.
Table~1 of Ref.~\cite{Albrecht:2026mdp} provides a convenient summary of typical reconstructed observables and dominant backgrounds 
in LHCb or Belle~II-type LFUV analyses.
In practice, binned fits in the marginal distributions of these can provide better control of systematics, 
and are thought to provide better overall discrimination between NP models.
On top of this, the abovementioned polarization and angular observables can be expected to play an indispensable role in discriminating between SM and NP, or different NP scenarios.
The optimal representation of the differential information is likely a combination of LFUV ratios, polarization, angular, or other shape observables, 
plus binned marginal kinematic distributions.

Following are open questions, in our view, that should enter into selection of observables 
optimized for NP sensitivity in the upcoming precision era:
\begin{enumerate}
	\item How do we balance the experimental systematic cleanliness of shape information or ratios versus statistical power in the full datasets including normalizations?
	\item Exclusive modes can be insensitive to particular NP WCs. E.g. by definition  $\wc{S}$ in $\Bbar \to \Dstar\tau\nu$, 
	and in practice $\wc{T}$ in $\Bbar \to X_c \ell \nu$~\cite{Carvunis:2025vab}.
	With a combination of modes, accounting for systematic correlations, what degree of discrimination power can be achieved between simplified models?
	\item Backgrounds such as $\Bbar \to D^{**} \tau \nu$ in $\Bbar \to \Dx \tau \nu$ analyses also are sensitive to NP,
	but in a characteristically different manner.
	Are there optimal selections in phase space, or shape observables, that leverage NP effects in both backgrounds and signal?
	\item For all of the above, how much does NP or hadronic template dependence matter?
\end{enumerate}

\section{Experimental status and strategies}

\subsection{Updates}
The experimental landscape of $R(\Dx)$ measurements comprises increasingly
precise results from BaBar~\cite{BaBar:2012obs,BaBar:2013mob},
Belle~\cite{Belle:2015qfa,Belle:2016dyj,Belle:2017ilt,Belle:2019rba},
LHCb~\cite{LHCb:2023uiv,LHCb:2023zxo,LHCb:2024jll}, and
Belle~II~\cite{Belle-II:2025yjp}.
The world average remains above the SM expectation for both $\RD$ and $\RDs$.

Hadron-collider experiments have extended these tests to other $b$-hadron species and final
states, exploiting the large $b$-hadron production cross section and integrated luminosity at the
LHC. LHCb has measured the ratio $\RJ$~\cite{LHCb:2017vlu}, the baryonic ratio
$\RLc$~\cite{LHCb:2022piu}, and excited charm states through $R(D^{**})$~\cite{LHCb:2025fri}.
CMS has also measured $\RJ$, using both leptonic and three-prong hadronic $\tau$
decays~\cite{CMS:2024seh,CMS:2025jfx}, providing the first contribution to these tests from
a general-purpose detector.
These ratios probe the same short-distance $\bctn$ transition in different hadronic environments,
with largely uncorrelated hadronic uncertainties.
Belle~II offers a complementary handle. In its clean $e^+e^-$ environment the initial state is
precisely known, and reconstructing the companion $B$ meson determines the momentum of the signal
$B$ and hence the total missing momentum.
This enables not only exclusive $R(\Dx)$ measurements but also a determination of the inclusive
ratio $\Rx{X}$~\cite{Belle-II:2023aih}.

All of these measurements share a common experimental difficulty: 
The semitauonic decay plus the subsequent $\tau$ decay produce at least two neutrinos. Unlike in the light-lepton normalization
mode, the missing mass, $\msqmiss$, is therefore not constrained to vanish, and signal and normalization events
cannot be separated event by event.
The separation is instead statistical, through template fits to kinematic distributions.
The precision of $R(\Dx)$ is consequently governed as much by the modeling of the signal,
normalization and background components as by the size of the available data sets.

Beyond these ratios, angular and polarization observables offer complementary sensitivity to NP
in $\bctn$ transitions, as discussed in Sec.~\ref{sec:optobs}.
The $\tau$ polarization $P_\tau(D^*)$ and the longitudinal $D^*$ polarization fraction
$F_{L,\tau}(D^*)$ are particularly powerful in discriminating between scenarios with scalar or tensor
operators. Such scenarios can reproduce the same enhancement of the integrated rate while
predicting distinctly different polarizations~\cite{Asadi:2018sym,Asadi:2019xrc}.
The first measurement of $P_\tau(D^*)$ was performed by Belle using hadronic $\tau$
decays~\cite{Belle:2016dyj,Belle:2017ilt}, while $F_{L,\tau}(D^*)$ has been measured by both
Belle~\cite{Belle:2019rba} and LHCb~\cite{LHCb:2023zxo}.
Because these observables are normalized ratios of yields, many hadronic and experimental
uncertainties cancel.
Should the tension in $R(\Dx)$ be established at high significance, they will likely be
indispensable for distinguishing between NP explanations.

\textbf{Experimental updates:}
Prior reviews have provided comprehensive surveys of the experimental landscape up to 2022~\cite{Bernlochner:2021vlv}. 
We summarize here the developments in
$\bctn$  LFUV measurements post-2022, 
covering both the new measurements themselves and the progress in the theoretical description of the dominant backgrounds that limit them.

Significant recent progress has come from LHCb, 
which has published three new determinations of $\RD$ and $\RDs$. 
The first of these was a measurement of $\Rx{D^*}=0.281(18)(24)$ and $\Rx{D^0}=0.441(60)(66)$, using the muonic $\tau^+ \to \mu^+ \nu_\mu \bar{\nu}_\tau$ channel with the Run~1 dataset~\cite{LHCb:2023zxo}. This was followed by  $\Rx{D^+}=0.249(43)(47)$ and $\Rx{D^{*+}}=0.402(81)(85)$ from partial Run~2 data ($2\,\mathrm{fb}^{-1}$), also with muonic $\tau$ decays~\cite{LHCb:2024jll}. The analysis in the three-prong channel $\tau^+ \to \pi^+\pi^-\pi^+(\pi^0)\bar{\nu}_\tau$, normalized to $B^0 \to D^{*-}\pi^+\pi^-\pi^+$, again using partial Run~2 data, measured $\Rx{D^{*+}}=0.260(15)(16)(12)$~\cite{LHCb:2023uiv}. In addition, LHCb recently reported their first simultaneous analysis of $R(D^0)=0.351(89)(72)(25)$ and $R(D^{0*})=0.258(26)(23)(20)$ with hadronic $\tau$ decays~\cite{LHCb:2026temp}.
These analyses exploit different $\tau$ reconstruction strategies with largely independent background compositions: 
the hadronic mode benefits from the reconstructible $\tau$ decay vertex and hence better $q^2$ resolution, 
but inherits the uncertainty of the external normalization branching fractions.
By contrast, the muonic mode avoids the latter uncertainty, but at the price of a degraded kinematic reconstruction. 

In parallel, Belle~II has produced first results based on its Run~1 data set. 
A first measurement of $\RDs$ with hadronic $B$ tagging and leptonic $\tau$ decays~\cite{Belle-II:2024ami} 
was superseded by a simultaneous determination  on the full $365\,\mathrm{fb}^{-1}$ Run~1 sample~\cite{Belle-II:2026nmn}, improving precision by roughly a factor of two, and with reported values $\RDs = 0.242(19)(16)$, $\RD = 0.439(55)(46)$. Additionally, Belle~II reported $\Rx{D^+}=0.418(74)(53)$ and $\Rx{D^{*+}}=0.306(34)(17)$ using semileptonic tagging and leptonic $\tau$ decays~\cite{Belle-II:2025yjp}. Belle~II also presented their first measurement on the hadronic $\tau$ final state analysis, measuring observables $R(D^*)=0.222(41)(30)$ and $P_\tau(D^*)=-0.33(1.15)(0.65)$~\cite{Belle-II:2026temp}. BaBar also presented a preliminary result based on semileptonic tag-side reconstruction and finds $R(D)=0.316(62)(19)$ and $R(D^*)=0.226(22)(12)$~\cite{Babar:RDstarSTApreliminary}.
These analyses demonstrate that the $B$-factory approach remains statistically competitive with LHC experiments.
For both LHCb and Belle~II analyses, systematic uncertainties are dominated by $B \to D^{**} \ell \nu$ background modelling and by the size of the simulated MC samples. 

A qualitatively new class of LFUV test is the first direct measurement of the inclusive ratio $\Rx{X}$ by Belle~II~\cite{Belle-II:2023aih}.
This measurement uses hadronically tagged events with a two-dimensional fit to $\msqmiss$ and the lepton momentum in the $B$ rest frame, $\pstarell$. 
As the inclusive rate predictions are (mostly) independent from the theoretical descriptions of the exclusive FFs, 
this observable is theoretically and experimentally decorrelated from $\RDx$.
It further includes the $14$--$20\%$ of the $\bctn$ rate that is not saturated by the sum of the $\Dx$ exclusive branching ratios 
per the Heavy Flavor Averaging Group HFLAV~\cite{HeavyFlavorAveragingGroupHFLAV:2024ctg}.

The doubly heavy ratio $\RJ$ has now been measured by CMS in the $J/\psi \to \mu^+\mu^-$, $\tau^+ \to \mu^+ \nu_\mu \bar{\nu}_\tau$~\cite{CMS:2024seh} and $\tau^+ \to 3\pi\bar{\nu}_\tau$~\cite{CMS:2025jfx} final state, obtaining $\RJ = 0.49(26)$,
and by LHCb with the Run~2~\cite{LHCb:2026dev} sample obtaining $\RJ = 0.51(12)(8)$. 
The resulting average by HFLAV of these measurements, together with the LHCb Run~1 analysis~\cite{LHCb:2017vlu}, is $\RJ = 0.54(12)$.
The CMS result has a large uncertainty, but is consistent with the LQCD SM prediction $\RJ = 0.2582(38)$~\cite{Harrison:2020nrv}, 
while the LHCb result agrees at the $1.7\sigma$ level. 
LHCb has also measured the LFUV ratio for the sum of the two narrow excited charm states, $R(D^{**0}_{1, 2})=0.13(3)(1)(2)$ with Run~1 and~2 data~\cite{LHCb:2025fri}.
No update to the baryonic ratio $\RLc$~\cite{LHCb:2022piu} has appeared to date.

Finally, LHCb has measured $F_{L,\tau}(D^*)$ in $B^0 \to D^{*-}\tau^+\nu_\tau$ with 3-prong $\tau$ decays
in two $q^2$ bins~\cite{LHCb:2023ssl}, 
obtaining a $q^2$-averaged value $F_{L,\tau}(\Dstar) = 0.41(6)(3)$ compatible with SM predictions $\simeq 0.46(6)$~\cite{Bernlochner:2021vlv}.
This new result has smaller uncertainties compared to Belle's earlier inclusive-tag determination~\cite{Belle:2019ewo}. 
The $\tau$ polarization $P_\tau(D^*)$ has not been remeasured since Refs.~\cite{Belle:2016dyj,Belle:2017ilt}, 
and remains the least constrained of the angular observables.

\textbf{Advancements on the $D^{**}$ systematics:}
A leading background and source of systematic uncertainty in $R(\Dx)$ measurements arises 
from theoretical descriptions of $\Bbar \to (\Dx \pi) l \nu$ decays, and in particular, 
the contribution from resonant $\Bbar \to D^{**} \to \Dx \pi$ transitions. 
This ``feed-down'' background arises when one pion is outside the detector acceptance (or otherwise misreconstructed),
such that background contributions mimic the signal semitauonic or normalization semileptonic modes.
In the following we focus on the $D\pi$ final state, keeping in mind similar considerations apply for $D^* \pi$.

Feed-down contributions have been described using an HQET-based parameterization~\cite{Leibovich:1997tu,Leibovich:1997em},
extended to include $\mathcal{O}(\alpha_s)$ and $\Lqcd/m_{c,b}$ corrections~\cite{Bernlochner:2016bci,Bernlochner:2017jka}.
This approach assumes implicitly a narrow-width approximation for the resonance, 
such that the $(D\pi)$ $S$-wave background is generated by a single broad $D_0^*(2300)$ resonance with Breit-Wigner lineshape,
and a $(D\pi)$ $D$-wave background is generated by a narrower $D_2^*(2460)$.
Because the shape of the resulting templates in the discriminating variables, typically 
 $\msqmiss$ and $E_\ell^*$ at $B$-factories, or the transverse variables employed by LHCb, is 
directly inherited from the assumed $q^2$ and $M_{D\pi}$ dependence of these decays, 
any mismodeling of the resonance composition propagates without suppression into the extracted signal yields. 
Together with the $\mathcal{O}(20\%)$ uncertainties on the light leptonic $\Bbar \to D^{**} \ell \nu$ branching fractions themselves, 
and the near-complete absence of experimental constraints on their tauonic counterparts, 
this renders the $D^{**}$ lineshape and partial wave description one of the leading systematic uncertainties in
the most recent $\RDx$ and $\Rx{X}$ analyses~\cite{LHCb:2023uiv,LHCb:2023zxo,Belle-II:2023aih}.

Deviations from the narrow-width approximation scale as $\Gamma/m$ or $(M_{D\pi}^2 - m^2)/m^2$, 
for the resonance pole mass $m$ and width $\Gamma$.
The relatively large width of the $D_0^*(2300)$, $\Gamma \gtrsim 200$\,MeV, means that such effects can be substantial,
while the virtuality of the resonance means that the FF formalism itself---based on on-shell selection rules---also requires modification.
Recent theoretical frameworks to account for such effects include an ad hoc partial-wave approach~\cite{Gustafson:2023lrz},
 partonic-level on-shell recursion relations plus HQET~\cite{Manzari:2024nxr} or heavy hadron chiral perturbation theory plus HQET~\cite{Papucci:2024qbt}.
Recent theoretical work further suggests that the single-resonance picture itself is inadequate.
Application of unitarized chiral perturbation theory predicts that the $D_0^*$ state 
contains two poles at $\simeq2.1\,\mathrm{GeV}$ and $\simeq 2.45\,\mathrm{GeV}$~\cite{Albaladejo:2016lbb,Du:2017zvv}: 
this is supported by LQCD results~\cite{Moir:2016srx,Gayer:2021xzv} 
and by a reinterpretation~\cite{Du:2020pui} of the LHCb amplitude analysis of $B^+ \to D^-\pi^+\pi^+$ decays~\cite{LHCb:2016lxy}. 
In addition, effects of the off-shell tail of the $D^*$ itself may be sizable~\cite{LeYaouanc:2018zyo,LeYaouanc:2021xcq}. 
Consequently, the same $M_{D\pi}$ spectrum can be accommodated by markedly different admixtures of $D_0^*$ $S$-wave, 
virtual $\Dstar$ $P$-wave, and $D_2^*$ $D$-wave background components.

A fit of the $M_{D\pi}$ spectrum measured by Belle~\cite{Belle:2022yzd} provided the first confrontation of $D^{**}$ background admixture hypotheses versus data.
This yielded a smaller $S$-wave and larger $D$-wave than typically assumed. 
The resulting $q^2$ spectrum was also significantly harder than in the HQET-based description, 
which shifts the feed-down background towards the kinematic region in which the $\tau$ signal is extracted.
Both effects have implications for precision $\RDx$ measurements.
Resolving the resonance composition experimentally, rather than inferring it from a fit to the invariant-mass spectrum alone, is the natural next step.
Angular or shape observables in the $\Bbar \to D \pi \ell \nu$ system may be particularly discriminating~\cite{Du:2025beb},
while theoretical inputs from HQET-based lineshape descriptions~\cite{Manzari:2024nxr,Papucci:2024qbt} may also enable systematically improvable methods.
The ultimate goal would be to replace largely model-driven $D^{**}$ systematic uncertainties by data-driven ones, 
via a full partial-wave analysis of $\Bbar \to \Dx \pi \ell \bar{\nu}_\ell$ decays at Belle~II.
Such control over $D^{**}$ backgrounds will be critical to enabling the next generation of precision $\RDx$ measurements.

\subsection{Global averaging and NP inference}
\label{sec:worldav}

The experimental world averages of $\RD$ and $\RDs$ compiled by HFLAV~\cite{HeavyFlavorAveragingGroupHFLAV:2024ctg} 
are generally accepted as authoritative, and taken by many phenomenological studies as the experimental inputs. 
The latest preliminary global average performed by HFLAV finds $\Rx{D^{0}} = 0.358(24)$ and $\Rx{D^{*0}} = 0.278(10)$~\cite{LHCb:2026temp}. The $\RDx$ combined deviation from the SM is currently $3.4\sigma$. 
It is therefore worth asking precisely what these averages mean,
and under which conditions they remain a faithful summary of the underlying data. 
Three challenges arise in this respect, becoming more severe as the measurements improve.

\textbf{The averaging procedure:}
The HFLAV combination is a Gaussian average of published central values and covariance
matrices, with correlations assigned analysis by analysis. 
This is adequate as long as the individual measurements are statistically dominated, 
because the likelihood of each input is then close to quadratic and fully characterized by its first two moments. 
This regime is, however, ending. 
Several $\RDx$ analyses are already systematics limited, 
and total uncertainties at the few-percent level are anticipated within the next decade~\cite{Bernlochner:2021vlv}.
The dominant systematic uncertainties in these analyses
are not from normalizations but \emph{shapes}: FF parameters,
$D^{**}$ feed-downs, and simulated-sample statistics all deform the templates in
$\msqmiss$, $q^2$ and $E_{\rm extra}$ in ways that are correlated bin-to-bin and, crucially,
across experiments. A analysis-wise covariance matrix on the extracted observables cannot encode
such a deformation, because it only records how much a nuisance parameter affects the result in each
analysis, and not on the templates themselves. Two analyses that involve the same FF parameter 
but use different fit observables will respond to a shift of that parameter in ways that no scalar correlation coefficient can capture.

\textbf{SM biases in NP inference:}
A more fundamental limitation concerns the meaning of experimental determinations of $\RDx$ once they are used for NP inference. 
Interpreting $\bctn$ data in the WET, one exploits the fact that the NP WCs modify not only the total rate but also the
differential decay. Precisely this shape sensitivity, however, undermines straightforward NP inference~\cite{Bernlochner:2020tfi}.
Most $\RDx$ measurements extract signal yields from MC template fits 
in which the signal, normalization and background shapes are generated under SM assumptions. 
If NP is present, it alters the kinematic distributions and thus the acceptances, 
resolutions and the reconstructed template shapes.
The $\RDx$ value recovered by an SM-based fit is then not the value that would be recovered under an NP hypothesis.
Ref.~\cite{Bernlochner:2020tfi} quantifies this with the \textsc{Hammer} reweighting library
by e.g. fitting SM templates to Asimov data sets generated at several NP truth points, 
finding that the recovered $\RDx$ values exclude the truth at the $3$--$4\sigma$ level 
for a data set corresponding to $5\,\mathrm{ab}^{-1}$ at Belle~II.
The recovered deviations are comparable in size to the present world average, and in some cases
point in a different direction from truth. 
Allowed and excluded regions of NP WC space derived by simplify fitting NP predictions to SM-based extractions of $\Rx{\Hc}$ (or other observables)
must therefore be read with corresponding caution, 
and motivates a forward-folding approach to NP inference (see Sec.~\ref{sec:optobs}).

\textbf{Shared hadronic effects in combinations:}
The third issue involves combinations.
The WCs $\wc{X}$ are universal to all measurements, but the hadronic matrix elements are not:
the $\Bbar \to \Dx$ FF parameters, for example, enter analyses as signal, normalization or background
and are fully correlated across experiments.
A first sensitivity study of a likelihood-level combination of LHCb-like and Belle~II-like
$\Bbar \to \Dx\tau\nu$ data~\cite{Albrecht:2026mdp} compares a simultaneous multi-channel fit,
in which the $\wc{X}$ and the FF parameters are shared, with a post-fit sum of independently
profiled single-channel likelihoods, which is close to how combinations are performed today.
Simulated events are reweighted to any point in WC and FF parameter space, so both are floated
continuously and propagated consistently through acceptance, resolution and template shapes,
without regenerating the detector simulation.

Because $\Bbar \to D$ and $\Bbar \to \Dstar$ have different degeneracy directions in WC space, 
the joint fit constrains the NP couplings considerably better than a naive combination of the individual results. 
Sharing the FF parameters tightens the constraints further, as channels with strong normalization handles fix hadronic uncertainties 
that could otherwise mimic NP elsewhere. Most importantly, if channels use inconsistent hadronic parameterizations, 
e.g.\ BGL for hadronic-$\tau$ and HQET-based for light-leptonic modes, 
the post-fit sum of SM-like data develops a spurious, displaced minimum in some NP WCs, 
even though each channel is well modeled on its own: the summed likelihoods correspond to no single consistent model, 
so residual mismodelling is absorbed differently in each channel and can mimic NP.

\textbf{Implications:}
A joint likelihood-level combination is not merely
more sensitive than a post-fit average, but is required for self-consistent NP inference on $\bctn$ data. 
Detector- and environment-specific uncertainties, 
trigger and reconstruction at LHCb, and tag-side efficiency and missing-energy modelling at Belle~II are largely 
uncorrelated and can be combined straightforwardly once likelihoods are available.
By contrast, the theory-driven inputs are logically shared and belong in a common nuisance-parameter set. 

Two routes lead to such a combination.
The first is for experiments to publish portable likelihoods in which the NP WCs and the leading
FF parameters appear explicitly, rather than summaries of observables obtained under SM assumptions.
The model-agnostic likelihood of the Belle~II $B^+ \to K^+\nu\bar{\nu}$
analysis~\cite{Belle-II:2025lfq,Gartner:2024muk} provides a template.
The second is to combine at the level of the analysis $n$-tuples, fitting the events of all
experiments in a single model.
This removes the summarization step entirely: shared FF parameters act identically in every channel,
no approximation of the input likelihoods is needed, and nobody has to anticipate at publication
time which nuisance parameters a future combination will want to share.
The obstacles are organizational rather than technical, including data preservation, blinding and publication
policies, and collaboration-specific calibration expertise.

\subsection{New physics and hadronic model degeneracies}
\label{sec:np-vs-hadronic}

\textbf{Wilson coefficients versus  form factors:}
Every term in a semileptonic amplitude involves a product of WCs and FFs. 
Thus a change in a NP WC may potentially be degenerate with a corresponding change in a FF parameter, unless: 
\begin{enumerate*}[label = \bf(\roman*)] 
\item The NP current alters kinematic distributions in an orthogonal direction to those generated by variations in FF parameters; or
\item Auxiliary information from LQCD inputs, continuum constraints (such as HQET or other theory frameworks; see Sec.~\ref{sec:thframe}) or light lepton modes constrains FFs independently.
\end{enumerate*}
The degree of potential degeneracy depends on the current. 
For instance, vector and axial-vector NP currents share the same FFs as the SM, some of which are constrained directly by light-lepton modes.
By contrast the tensor FFs enter multiplicatively with $\wc{T}$, and hence only their product is constrained, yielding an exact degeneracy between the WC and the tensor FF parameters.
An inference about the NP WCs contributing to the given process is thus only as good as the knowledge of the FFs that accompany them, 
and that knowledge has to be precise compared with the NP effect being analyzed.

\textbf{Current LQCD tensions in the normalization mode:}
As outlined in Sec.~\ref{sec:thframe} the normalization of the $\Bbar \to \Dstar$ tensor FFs is known to about $2\%$,
while higher order BGL coefficients carry uncertainties of $20$--$90\%$. 
However, constraining the FFs with LQCD is already problematic for the light $\Bbar \to \Dstar\ell\nu$ normalization mode, 
assumed to be SM.
Combined analyses of the Belle and Belle~II differential distributions with the LQCD results find tensions in the shapes, 
both between lattice predictions and data and among different lattice inputs~\cite{Bordone:2024weh,Fedele:2023ewe}.
Simulated $\Bbar \to \Dstar\ell\nu$
samples generated with lattice FFs therefore do not reproduce the measured shapes.
This leaves two options for NP inference using the $\tau$ mode:
\begin{enumerate*}[label = \bf(\roman*)] 
\item Templates built with LQCD FFs inherit a shape mismodelling that is present even where no NP is expected, and this is absorbed into the inferred WC CLs;
\item FFs fitted to the light-lepton data describe the SM FFs empirically,
	but leave $\tau$-specific FFs---the pseudoscalar `$P_1$' and tensor FFs---unconstrained, 
	unless one can tie them to the modelling of the normalization mode with other theory inputs.
\end{enumerate*}
Until this tension is understood, NP inference of $\Bbar \to \Dstar\tau\nu$ data is limited in its use of LQCD inputs. 

\begin{figure}[t]	
	\centering
    \includegraphics[width=\linewidth]{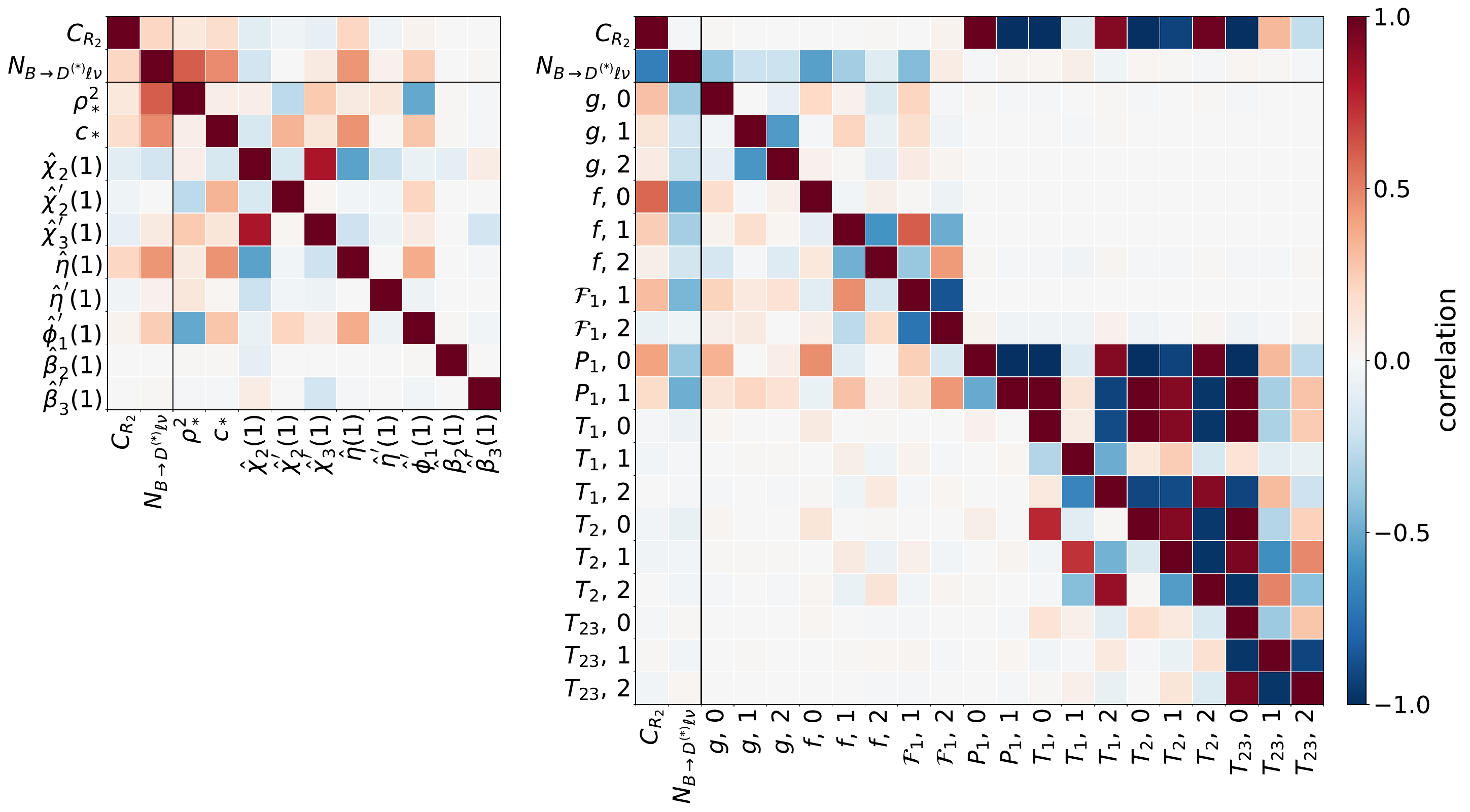}
	\vspace{-10pt}
  \caption{Results of an Asimov data study of Belle~II $B\to D^*\tau\nu$ pseudo-data ($5\,\mathrm{ab}^{-1}$) for the $R_2$ leptoquark, $\wc{$R_2$} = \wc{SL} = 0.25 \simeq 8\wc{T}$.
  \textbf{Left:} Correlations of $\wc{$R_2$}$  with the $B\to D^*\ell\nu$
  normalization $N_{B\to D^{(*)}\ell\nu}$ and the 10 BLPRXP FF parameters~\cite{Bernlochner:2022ywh} per FF scenario~\ref{sc:BLPRXP} in the main text.
  \textbf{Right:} Correlations of $\wc{$R_2$}$ with $N_{B\to D^{(*)}\ell\nu}$ and 19 BGL coefficients per FF scenario~\ref{sc:BGL}.
  Each parameter BGL $a_n^\text{FF}$ is labelled  as ``$\text{FF}, n$''.
  In the lower triangle, both the 10 SM BGL parameters and the 9 BGL tensor parameters ($a^{T_1}_n$, $a^{T_2}_n$, $a^{T_{23}}_n$) are constrained by LQCD; in the upper triangle, ($a^{P_1}_n$, $a^{T_1}_n$, $a^{T_2}_n$, $a^{T_{23}}_n$) are unconstrained.}
  \label{fig:R2corr}
\end{figure}

\textbf{Example:} The degeneracies can be seen in an Asimov study 
of Belle~II-like data for $\Bbar \to \Dx \tau\nu$ ($\tau\to\ell\nu\bar\nu$, $5\,\mathrm{ab}^{-1}$), 
fitted in $\pstarell$ and $\msqmiss$
using \texttt{Hammer}~\texttt{v2.0} \cite{Bernlochner:2020tfi, bernlochner_2026_20716678,Papucci:2026qhd}.
Using the approach of Ref.~\cite{Bernlochner:2020tfi} we consider two FF scenarios: 
\begin{enumerate*}[label = \bf(\Alph*)] 
\item \label{sc:BLPRXP} The HQET-based ``BLPRXP'' parametrization~\cite{Bernlochner:2022ywh}, 
which includes second-order HQET corrections with floating IW parameters. 
This approach involves a combined fit of $\Bbar \to \Dstar l \nu$ and $\Bbar \to D l \nu$ data
and includes (beyond) zero-recoil LQCD data for $\Bbar \to \Dstar \ell \nu$ ($\Bbar \to D \ell \nu)$;
\item \label{sc:BGL} The dispersive BGL parametrization~\cite{Boyd:1997kz}, 
floating all SM and NP BGL parameters, denoted $a^{FF}_k$, including tensor BGL parameters~\cite{Bordone:2025jur}.
The SM parameters in $\Bbar \to \Dstar \ell \nu$ are constrained by 
a BGL fit to Belle data~\cite{Bernlochner:2019ldg} for $g$, $f$, $\mathcal{F}_1$.
The coefficients of $P_1$ and the NP tensor FFs are either all left free or all constrained by LQCD~\cite{Harrison:2023dzh}.
\end{enumerate*}

We generate Asimov $\Bbar \to \Dx \tau \nu$ data for the $R_2$ leptoquark simplified model with operator $\op{SL} + \op{T}/(4r)$ such that the corresponding WC $\wc{$R_2$} = \wc{SL} = 0.25 \simeq 8\wc{T}$. 
The $\Dstar$ category has 4750 $\Bbar \to \Dstar \tau \nu$, 82500 $\Bbar \to D \ell \nu$ and 1440 background events. 
The $D$ category (BLPRXP only) has 4000 $\Bbar \to D \tau \nu$, 1625 $\Bbar \to \Dstar \tau \nu$ feed-down, 
35000 $\Bbar \to D \ell \nu$, 28750 $\Bbar \to \Dstar \ell \nu$ feed-down and 1440 background events. 
The background is treated as flat and fixed. 
To approximate detector effects: we use the Belle~II polar acceptance; minimum lepton energy is 300 MeV;
the slow-pion efficiency has a turn-on starting at 40 MeV; a Gaussian $0.5$ GeV resolution is applied on $\msqmiss$; $\pstarell$ is not smeared.

We show in Fig.~\ref{fig:R2corr} the correlations between the NP WCs and the FF parameters for both FF scenarios~\ref{sc:BLPRXP} and~\ref{sc:BGL},
noting that the latter has two variants depending on whether or not LQCD inputs for the NP $\Bbar \to \Dstar \tau \nu$ modes are used.
In scenario~\ref{sc:BLPRXP}, the HQET relations between the SM and NP FFs decorrelate the NP WCs from the FF parameters almost completely.
NP inference is then not limited by knowledge of the FFs:
The achievable statistical precision on $C_{R_2}$ is about $5\%$, with correlations below $0.2$ to any other fit parameter.
By contrast, in scenario~\ref{sc:BGL}, the result depends on whether LQCD constraints are included on the pseudoscalar and tensor coefficients $a^{P_1}_k$ and $a^{T_i}_k$ in $\Bbar \to \Dstar \tau \nu$.
With the LQCD constraints included, $C_{R_2}$ is measured to about $15\%$ precision, and it correlates or anticorrelates
mainly with the light-lepton normalization and $a^f_0$, and to a lesser extent with $a^{P1}_0$.
Without them, contributions to the NP signal from scalar and tensor WCs are fully absorbed by the $a^{P_1}_k$ and $a^{T_i}_k$ FF coefficients, 
and $C_{R_2}$ can no longer be precisely measured.
A SM template then describes the data almost as well as the $R_2$ model, 
and the discovery significance is lowered from about $7\sigma$ to about $3\sigma$ in this example.

While this example uses Asimov data, a fast simulation and a single benchmark, it nonetheless illustrates the following lessons:
\begin{enumerate*}[label = \bf(\roman*)] 
\item Obtaining CLs on the NP WCs, especially the pseudoscalar and tensor NP WCs, without degeneracies generically requires hadronic uncertainties well below the size of NP signal.
This can be achieved by LQCD inputs, or HQET relations, or other model-dependent inputs.
\item The $\tau$ and light-lepton modes should be fitted simultaneously with common FF parameters, 
with the light-lepton normalization should be constrained by auxiliary data, noting the WCs may inherit any tensions from the latter;
\item WC CLs should be published with their correlations to the FF parameters and to the light-lepton normalization. 
They should be checked for stability under alternative FF inputs (LQCD only, LQCD plus experiment, each LQCD calculation separately);
\item It is crucial for experiments to properly characterize whether their measured distributions can distinguish between contributions from NP WCs and hadronic effects.
\end{enumerate*}

\subsection{Outlook: beyond semileptonics}

Precision measurement of $\bctn$ semileptonic decays and the establishment of $\Rx{\Hc}$ or $\Rx{X_c}$ 
LFUV anomalies in these charged-current processes may be the discovery channel for new physics.
In this review, we have provided a comprehensive survey of the central theoretical concepts and experimental considerations 
that must be mastered and brought under control for self-consistent and unbiased NP interpretations of the LFUV anomalies.
We have examined the status and challenges to present-day and future precision analyses, 
as well as promising strategies to enable the percent-level era of precision that may be required to conclusively establish or reject new physics.

While $\bctn$ decays may be the discovery channel, 
as a parting note we emphasize that explanations of the LFUV anomalies must not be assessed in isolation:
measurements across a range of channels may be required to conclusively understand what NP is at play.
As discussed in Sec.~\ref{sec:npsl}, EW gauge invariance requires the operators that modify
$\bctn$ to be built from the third-generation quark and lepton
doublets, that may also induce neutral-current transitions
such as $b\to s\nu\bar\nu$,
$b\to s\tau^+\tau^-$, and the flavour-diagonal $b\bar b\to\tau^+\tau^-$,
as well as $\Delta B=2$ transitions, in particular $B_s$--$\bar B_s$ mixing.
As some examples: the $U_1$ leptoquark cannot generate $b\to s\nu\bar\nu$ at tree level, 
but predicts enhancements of $B_s\to\tau^+\tau^-$ and $B\to K\tau^+\tau^-$~\cite{Cornella:2021sby}, 
while $S_1$ and $S_3$ generically contribute to $B\to K^{(*)}\nu\bar\nu$, where Belle~II has reported evidence for an enhanced rate~\cite{Belle-II:2023esi}.
The $R_2$ coupling is constrained by the $B_c$ lifetime and by high-$p_T$ searches. 
$B_s$--$\bar B_s$ mixing is generated at one loop in leptoquark models, but at tree level by the $Z'$ partner of a $W_3'$~\cite{Greljo:2015mma,Buttazzo:2017ixm}, 
which limits the effective flavor violating $b$--$s$ coupling. 
Finally, without light right-handed neutrinos, new physics in
$\bctn$ implies effects in $b\bar b\to\tau^+\tau^-$~\cite{Aloni:2017eny}, 
which are testable in $\Upsilon(nS)\to\tau^+\tau^-$ decays and, at higher energies, in
$pp\to\tau^+\tau^-$~\cite{Faroughy:2016osc} at the LHC. 

Thus in a bright future that NP is established in $\bctn$ decays, 
a rich experimental program at Belle~II and the LHC will be required to draw together a wide range of precision measurements 
into a self-consistent combination and identification of phenomenology beyond the Standard Model.

\paragraph{Acknowledgments}
We thank Pouya Asadi and Marcello Rotondo for comments on the manuscript.
M.P. is supported by the German Research Foundation (DFG) Emmy-Noether Grant No. 526218088.
D.J.R. is supported by the Office of High Energy Physics of the U.S. Department of Energy under contract DE-AC02-05CH11231.
AI assisted coding was used for the MC generation in the Asimov study in Sec.~\ref{sec:np-vs-hadronic} and in the generation of Fig~\ref{fig:R2corr}.
The manuscript text, tables, and other figures contain no generative AI content.

\input{AnnRev-LFU_arXiv_v1.bbl}

\end{document}

%% file: AnnRev-LFU_arXiv_v1.bbl
\ifx\mcitethebibliography\mciteundefinedmacro
\PackageError{unsrtM.bst}{mciteplus.sty has not been loaded}
{This bibstyle requires the use of the mciteplus package.}\fi
\def\enquote#1{``#1''}
\expandafter\ifx\csname url\endcsname\relax
  \def\url#1{{\tt #1}}\fi
\expandafter\ifx\csname urlprefix\endcsname\relax\def\urlprefix{URL }\fi
\expandafter\ifx\csname eprint\endcsname\relax\def\eprint#1{\url{#1}}\fi